\documentclass[11pt]{article}
\usepackage{amssymb}
\usepackage{epsfig}

\newcommand{\BE}{\begin{equation}}
\newcommand{\EE}{\end{equation}}

\begin{document}

\vspace*{1mm}
\begin{center}

\vskip 1 pt

  {\LARGE{\bf {The CMB, preferred reference system and dragging of light in the earth frame}}}

\end{center}

\begin{center}
\vspace*{14mm} {\Large  M. Consoli$^{(a)}$ and A. Pluchino
$^{(b,a)}$}
\vspace*{4mm}\\
{a) Istituto Nazionale di Fisica Nucleare, Sezione di Catania, Italy ~~~~~~~~~\\
b) Dipartimento di Fisica e Astronomia dell'Universit\`a di Catania,
Italy }
\end{center}

\par\noindent The dominant CMB dipole anisotropy is a Doppler effect due to a
particular motion of the solar system with velocity of 370 km/s.
Since this derives from peculiar motions and local inhomogeneities,
one could meaningfully consider a fundamental frame of rest $\Sigma$
associated with the Universe as a whole. From the group properties
of Lorentz transformations, two observers, individually moving
within $\Sigma$, would still be connected by the relativistic
composition rules. But ultimate implications could be substantial.
Physical interpretation is thus traditionally demanded to
correlating some dragging of light observed in laboratory with the
direct CMB observations. Today the small residuals, from
Michelson-Morley to present experiments  with optical resonators,
are just considered instrumental artifacts. However, if the velocity
of light in the interferometers is not the same parameter  ``c''  of
Lorentz transformations, nothing would prevent a non-zero dragging.
Furthermore, observable effects would be much smaller than
classically expected and most likely of irregular nature. We review
an alternative reading of experiments which leads to remarkable
correlations with the CMB observations. Notably, we explain the
irregular $10^{-15}$ fractional frequency shift presently measured
with optical resonators operating in vacuum and solid dielectrics.
For integration times of about 1 second, and typical Central-Europe
latitude, we also predict daily variations of the Allan variance in
the range $(5\div12) \cdot 10^{-16}$.



\section{Introduction}

Soon after the discovery \cite{penzias} of the Cosmic Microwave
Background (CMB) it was realized that the observed temperature of
the radiation should exhibit a small anisotropy as a consequence of
the Doppler effect associated with the motion of the earth
\cite{partridge,heer} ($\beta=V/c)$ \BE
T(\theta)={{T_o\sqrt{1-\beta^2}}\over{1- \beta \cos \theta} } \EE
Accurate observations with satellites in space \cite{mather,smoot}
have shown that the measured temperature variations correspond to a
motion of the solar system described by an average velocity $V\sim
370$ km/s, a right ascension $\alpha \sim 168^o$ and a declination
$\gamma\sim -7^o$, pointing approximately in the direction of the
constellation Leo. This means that, if one sets $T_o \sim $ 2.725 K
and $\beta\sim 0.00123$, there are angular variations of a few
millikelvin \BE \label{CBR}\Delta T^{\rm CMB}(\theta) \sim T_o \beta
\cos\theta \sim \pm 3.36 ~{\rm mK} \EE These variations represent by
far the largest contribution to the CMB anisotropy and are usually
denoted as the {\it kinematic dipole} \cite{yoon}.

With this interpretation, it is natural to wonder about the
reference frame where this CMB dipole vanishes exactly, i.e. could
it represent a fundamental system for relativity as in the original
Lorentzian formulation? The standard answer is that one should not
confuse these two concepts. The CMB is a definite medium and sets a
rest frame where the dipole anisotropy is zero. There is nothing
strange that our motion with respect to this system can be detected.
In this sense, there would be no contradiction with special
relativity.

Though, to good approximation, this kinematic dipole arises from the
vector combination of the various forms of peculiar motion which are
involved (rotation of the solar system around the center of the
Milky Way, motion of the Milky Way toward the center of the Local
Group, motion of the Local Group of galaxies in the direction of the
Great Attractor...) \cite{smoot}. Therefore, since the observed CMB
dipole reflects local inhomogeneities, it becomes natural to imagine
a global frame of rest associated with the Universe as a whole. The
isotropy of the CMB could then just {\it indicate} the existence of
this fundamental system $\Sigma$ that we may conventionally decide
to call ``ether'' but the cosmic radiation itself would not {\it
coincide} with this form of ether \footnote{With very few
exceptions, modern textbooks tend to give a negative meaning to the
idea of a fundamental state of rest. Yet, this was the natural
perspective for the first derivation of the relativistic effects by
Lorentz, Fitzgerald and Larmor. Over the years, the value of a
Lorentzian formulation has been emphasized by many authors, notably
by Bell \cite{bell}, see Brown's book \cite{brownbook} for a
complete list of references. For more recent work, see also De Abreu
and Guerra \cite{guerraejtp} and Shanahan \cite{shanahan}.}. Due to
the group properties of Lorentz transformations, two observers S'
and S'', moving individually with respect to $\Sigma$, would still
be connected by a Lorentz transformation with relative velocity
parameter fixed by the standard relativistic composition rule
\footnote{We ignore here the subtleties related to the Thomas-Wigner
spatial rotation which is introduced when considering two Lorentz
transformations along different directions, see e.g. \cite{ungar,
costella, kanevisser}.}. But, ultimate consequences could be far
reaching. Just think to the implications for the interpretation of
non-locality in the quantum theory \footnote{This was well
illustrated in ref.\cite{hardy}: ``Thus, Nonlocality is most
naturally incorporated into a theory in which there is a special
frame of reference. One possible candidate for this special frame of
reference is the one in which the cosmic background radiation is
isotropic. However, other than the fact that a realistic
interpretation of quantum mechanics requires a preferred frame and
the cosmic background radiation provides us with one, there is no
readily apparent reason why the two should be linked''.}.

The idea of a preferred frame finds further motivations in the
modern picture of the vacuum, intended as the lowest energy state of
the theory. This is not trivial emptiness but is believed to arise
from the macroscopic Bose condensation process of Higgs quanta,
quark-antiquark pairs, gluons... see e.g.
\cite{thooft,mech,epjc,dedicated,foop}. The hypothetical global
frame could then reflect a vacuum structure which has a certain
substantiality and can determine the type of relativity physically
realized in nature.

Since the answer cannot be found with theoretical arguments only,
the physical role of $\Sigma$ is thus traditionally postponed to the
experimental observation, in the earth frame $S'$, of some dragging
of light: the effect of an ``ether drift''. This would require: i)
to detect in laboratory a small angular dependence
${{\Delta\bar{c}_\theta}\over{c}} \neq 0$ of the two-way velocity of
light and ii) to correlate this angular dependence with the direct
CMB observations with satellites in space.

Of course, experimental evidence for both the undulatory and
corpuscular aspects of radiation has substantially modified the
consideration of an ether and its logical need for the physical
theory. Yet, the existence of a rest frame tight to the underlying
energy structure of the vacuum does not contradict the basic tenets
of general relativity where the off-diagonal components $g_{0i}$ of
the metric play the role of a velocity field and, as such, are the
most natural way to introduce effects associated with the state of
motion of the observer as, for instance, a small angular dependence
of the velocity of light \footnote{Preferred-frame effects are
common to many models of dark-energy (and/or of dark-matter) such as
the massive gravity scheme proposed by Rubakov \cite{Rubakov}, or
the effective graviton-Higgs mechanism of ref.\cite{arrault} or
non-local modifications of the Einstein-Hilbert action
\cite{deser,soussa,Nojiri}. In these cases, one also expects a
dependence of the velocity of light on the state of motion of the
observer.}.

So far, it is generally believed that no genuine ether drift has
ever been observed. In this traditional view, which dates back to
the end of XIX century, when one was still comparing with Maxwell's
classical predictions for the orbital velocity $v_{\rm orb}= $ 30
km/s, all measurements (from Michelson-Morley to the most recent
experiments with optical resonators) are seen as a long sequence of
null results, i.e. typical instrumental effects in experiments with
better and better systematics (see e.g. Figure 1 of
ref.\cite{nagelnature}).

However, to a closer look, things are not so simple for at least
three reasons:

~~~i) In the old experiments (Michelson-Morley, Miller, Tomaschek,
Kennedy, Illingworth, Piccard-Stahel, Michelson-Pease-Pearson, Joos)
\cite{mm}-\cite{joos}, light was propagating in gaseous media, air
or helium at room temperature and atmospheric pressure. In these
systems with refractive index ${\cal N}=1 + \epsilon$ the velocity
of light in the interferometers, say $c_\gamma$, is not the same
parameter $c$ of Lorentz transformations. Hence nothing prevents a
non-zero effect because, when light gets absorbed and re-emitted,
the small fraction of refracted light could keep track of the
velocity of matter with respect to the hypothetical $\Sigma$ and
produce a direction-dependent refractive index. Then, from symmetry
arguments valid in the $\epsilon \to 0$ limit
\cite{pla}-\cite{book}, one would expect
${{|\Delta\bar{c}_\theta|}\over{c}} \sim \epsilon (v^2/c^2)$ which
is much smaller than the classical expectation
${{|\Delta\bar{c}_\theta|_{\rm class}}\over{c}} \sim (v^2/2c^2)$.
For instance, in the old experiments in air (at room temperature and
atmospheric pressure where $\epsilon\sim 2.8\cdot 10^{-4}$) a
typical value was ${{|\Delta\bar{c}_\theta|_{\rm exp}}\over{c}} \sim
3\cdot 10^{-10}$. This was classically interpreted as a velocity of
7.3 km/s but would now correspond to 310 km/s. Analogously, in the
old experiment in gaseous helium (at room temperature and
atmospheric pressure, where $\epsilon\sim 3.3\cdot 10^{-5}$), a
typical value was ${{|\Delta\bar{c}_\theta|_{\rm exp}}\over{c}} \sim
2.2\cdot 10^{-11}$. This was classically interpreted as a velocity
of 2 km/s but would now correspond to 240 km/s. Those old
measurements could thus become consistent with the motion of the
earth in the CMB.

~~~ii) Differently from those old measurements, in modern
experiments light now propagates in a high vacuum or in solid
dielectrics, often in the cryogenic regime. Then, the present more
stringent limits might not depend on the technological progress only
but also on the media that are tested thus preventing a
straightforward comparison.

~~~iii) In the analysis of the data, the hypothetical signal of the
drift was always assumed a {\it regular } phenomenon, with only
smooth time modulations depending deterministically on the rotation
of the earth (and its orbital revolution). The data, instead, had
always an irregular behavior, with statistical averages much smaller
than the individual measurements, inducing to interpret the
measurements as typical instrumental artifacts. But a relation, if
any, between macroscopic motion of the earth and microscopic
propagation of light in laboratory depends on a complicated chain of
effects and, ultimately, on the nature of the physical vacuum. By
comparing with the motion of a body in a fluid, the traditional view
corresponds to a form of regular (``laminar'') flow where global and
local velocity fields coincide. Some general arguments, see
refs.\cite{chaos,physica}, suggest instead that the physical vacuum
might behave as a stochastic medium which resembles a turbulent
fluid where large-scale and small-scale flows are only related {\it
indirectly}. This means that the projection of the global velocity
field at the site of the experiment, say $ \tilde v_\mu(t)$, could
differ non trivially from the local field $v_\mu(t)$ which
determines the direction and magnitude of the drift in the plane of
the interferometer. Therefore, in the extreme limit of a turbulence
which becomes isotropic at the small scale of the experiment, a
genuine non-zero signal can coexist with vanishing statistical
averages for all vector quantities. In this perspective, one should
Fourier analyze the data for ${{\Delta\bar{c}_\theta(t)}\over{c}}$
and extract the (2nd-harmonic) phase $\theta_2(t)$ and amplitude
$A_2(t)$, which give respectively the direction and magnitude of the
effect, and concentrate on the amplitude which, being positive
definite, remains non-zero under any averaging procedure. By
correlating the local $v_\mu(t)$ with the global $ \tilde v_\mu(t)$,
the time modulations of the statistical average $\langle A_2(t)
\rangle_{\rm stat}$ can then give information on the magnitude,
right ascension and declination of the cosmic motion. Depending on
the type of correlation, there would be various implications. For
instance, in a simplest uniform-probability model, where the
kinematic parameters of the global $ \tilde v_\mu(t)$ are just used
to fix the typical boundaries for a local random $v_\mu(t)$, one
finds $\langle A_2(t) \rangle_{\rm stat}= (\pi^2/18) {\tilde A}_2(t)
$, where $\tilde A_2(t)$ is the amplitude in the deterministic
picture. With such smaller statistical average, one will obtain a
velocity larger by $\sqrt{18/\pi^2}\sim$ 1.35 from the same data.
Therefore, by returning to those old measurements
${{|\Delta\bar{c}_\theta|_{\rm exp}}\over{c}} \sim 3\cdot 10^{-10}$
and ${{|\Delta\bar{c}_\theta|_{\rm exp}}\over{c}} \sim 2.2\cdot
10^{-11}$, respectively for air or gaseous helium at atmospheric
pressure, the data can be interpreted in three different ways: a) as
7.3 and 2 km/s, in a classical picture b) as 310 and 240 km/s, in a
modern scheme and in a smooth picture of the drift c) as 418 and 324
km/s, in a modern scheme but now allowing for irregular fluctuations
of the signal. In this third interpretation, the average of the two
values agrees very well with the CMB velocity of 370 km/s.

After having illustrated why the evidences for $\Sigma$ may be much
more subtle than usually believed, we will review in Sect.2 the
basics of these experiments and, in Sects.3 and 4, the alternative
theoretical framework of refs.\cite{plus}-\cite{book}. This will be
applied in Sect.5 to the old experiments in gaseous media where
${{\Delta\bar{c}_\theta}\over{c}}$ was extracted from the fringe
shifts in Michelson interferometers. As we will show, our scheme,
which discards the phase and just focuses on the 2nd-harmonic
amplitudes, leads to a consistent description of the data and to
remarkable correlations with the direct CMB observations with
satellites in space.

As it often happens, symmetry arguments can successfully describe a
phenomenon regardless of the physical mechanisms behind it. The same
is true here with our relation
${{|\Delta\bar{c}_\theta|}\over{c}}\sim \epsilon (v^2/c^2)$. It
gives a consistent description of the data but does not explain how
the earth motion produces the tiny observed anisotropy in the
gaseous systems. To this end, as a first possibility, we have
considered that the electromagnetic field of the incoming light
could determine different polarizations in different directions in
the dielectric, depending on its state of motion. However, if this
works in weakly bound gaseous matter, the same mechanism should also
work in a strongly bound solid dielectric, where the refractivity is
$({\cal N}_{\rm solid} -1)= O(1)$, and thus produce a much larger
${{|\Delta\bar{c}_\theta|}\over{c}}\sim ({\cal N}_{\rm solid} -1)
(v^2/c^2)\sim 10^{-6} $. This is in contrast with the Shamir-Fox
\cite{fox} experiment in perspex where the observed value was
smaller by orders of magnitude. As an alternative possibility, we
have thus re-considered in Sect.6 the traditional thermal
interpretation \cite{joos2,shankland} of the observed residuals. The
idea was that, in a weakly bound system as a gas, a small
temperature difference $\Delta T^{\rm gas}(\theta)$, of a
millikelvin or so, in the air of the optical arms could produce
density changes and a difference in the refractive index
proportional to $\epsilon_{\rm gas} \Delta T^{\rm gas}(\theta)/T$,
where $T\sim$ 300 K is the temperature of the laboratory. Miller was
aware of this potentially large effect \cite{miller} and objected
that casual changes of temperature would largely cancel when
averaging over many measurements. Only temperature effects which had
a definite periodicity would survive. The overall consistency, in
our scheme, of different experiments would now indicate that such
$\Delta T^{\rm gas}(\theta)$ must have a {\it non-local} origin as
if, for instance, the interactions with the background radiation
could transfer a part of $\Delta T^{\rm CMB}(\theta)$ in
Eq.(\ref{CBR}) and bring the gas out of equilibrium. Only, those old
estimates were slightly too large because our analysis gives $\Delta
T^{\rm gas}(\theta)= (0.2\div 0.3)$ mK suggesting that the
interactions are so weak that, on average, the induced temperature
differences in the optical paths were only 1/10 of the $\Delta
T^{\rm CMB}(\theta)$ in Eq.(\ref{CBR}). Nevertheless, whatever its
precise value, this typical magnitude can help intuition. In fact,
it can explain the {\it quantitative} reduction of the effect in the
vacuum limit where $\epsilon_{\rm gas} \to 0$ and the {\it
qualitative} difference with solid dielectrics where such small
temperature differences cannot produce any appreciable deviation
from isotropy in the rest frame of the medium.

Most significantly, this thermal argument has also an interesting
predictive power. In fact, it implies that if a very small, but
non-zero, fundamental signal were definitely detected in vacuum
then, with very precise measurements, the same signal should also
show up in a solid dielectric where temperature differences of a
millikelvin or so become irrelevant. In Sect.7, this expectation
will be compared with the modern experiments where
${{\Delta\bar{c}_\theta}\over{c}}$ is now extracted from the
frequency shift of two optical resonators. Here, after the vector
average of many observations, the present limit is a residual
$\langle {{\Delta\bar{c}_\theta}\over{c}}\rangle =10^{-18}\div
10^{-19}$. However, this just reflects the very irregular nature of
the signal because its typical {\it magnitude}
${{|\Delta\bar{c}_\theta(t)|}\over{c}}\sim 10^{-15}$ is about 1000
times larger. This $10^{-15}$ magnitude is found with vacuum
resonators \cite{mueller2003}$-$\cite{schiller2015} made of
different materials, operating at room temperature and/or in the
cryogenic regime, and in the most precise experiment ever performed
in a solid dielectric \cite{nagelnature}. As such, it could hardly
be interpreted as a spurious effect. In the same model discussed
above, we are then lead to the concept of a refractive index ${\cal
N}_v$ for the physical vacuum which is established in an apparatus
placed on the earth surface. This ${\cal N}_v$ should differ from
unity at the $10^{-9}$ level, in order to give $
{{|\Delta\bar{c}_\theta(t)|_v}\over{c}}\sim ({\cal N}_v
-1)~(v^2(t)/c^2)~ \sim 10^{-15}$, and thus would fit with
ref.\cite{gerg} where a vacuum refractivity $\epsilon_v=({\cal N}_v
-1)\sim 10^{-9}$ was considered. Indeed, if the curvature observed
in a gravitational field reflects local deformations of the physical
space-time units, for an apparatus on the earth surface there might
be a tiny refractivity $\epsilon_v\sim (2G_NM/c^2R) \sim 1.4\cdot
10^{-9}$ where $G_N$ is the Newton constant and $M$ and $R$ the mass
and radius of the earth. This could make a difference with that
ideal free-fall environment which is always assumed to define
operationally the parameter $c$ of Lorentz transformations in the
presence of gravitational effects. Then, for a typical daily
projection 250 km/s $\lesssim \tilde v(t)\lesssim$ 370 km/s, and in
the same uniform-probability model used successfully for the
classical experiments, we would expect a fundamental signal with
average magnitude $ (8.5 \pm 3.5) \cdot 10^{-16}$. This is a genuine
signal which would pose an intrinsic limitation to the precision of
measurements and that, from our numerical simulations, can be
approximated as a white noise. Thus it should be compared with the
frequency shift of two optical resonators at the largest integration
time (typically 1 second) where the pure white-noise branch is as
small as possible but other types of noise are not yet important.

As emphasized in the conclusive Sect.8, the consistency of this
prediction with the most precise measurements in vacuum and solid
dielectrics, operating at room temperature and in the cryogenic
regime, and the satisfactory description of the old experiments
should therefore induce to perform an ultimate experimental check:
detecting the expected, periodic, daily variations in the range
$(5\div12) \cdot 10^{-16}$.

\section{Basics of the ether-drift experiments}

\begin{figure}
\begin{center}
\includegraphics[width=6.5 cm]{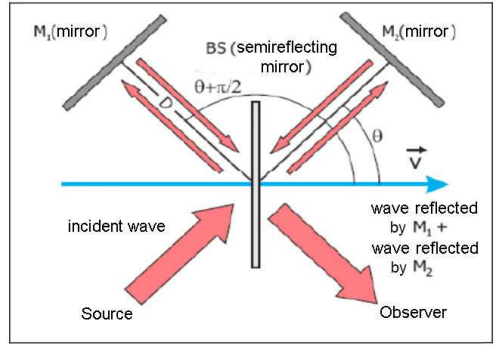}
\end{center}
\caption{A schematic illustration of the Michelson
interferometer.}
\label{Michinterferometer}
\end{figure}

Let us start with some basic notions. As anticipated, old and modern
experiments adopt a different technology but, in the end, have the
same scope: looking for the hypothetical $\Sigma$ through a tiny
angular dependence of the two-way velocity of light
$\bar{c}_\gamma(\theta)$. This quantity can be measured
unambiguously and is defined through the one-way velocity
$c_\gamma(\theta)$ as

\begin{equation}
\label{first0}
\bar{c}_\gamma(\theta)={{2~c_\gamma(\theta)c_\gamma(\pi
+\theta)}\over{c_\gamma(\theta) +c_\gamma (\pi +\theta)}}
\end{equation}
where $\theta$ indicates the angle between the direction where light
propagates and the velocity with respect to $\Sigma$. By defining
the anisotropy
\[ \Delta\bar{c}_\theta =
\bar{c}_\gamma(\pi/2+\theta)-\bar{c}_\gamma(\theta) \] one finds a
simple relation with  $\Delta t(\theta)$, the difference in time for
light propagating back and forth along perpendicular rods of length
$D$, see Fig.\ref{Michinterferometer}, and one finds
\begin{equation} \label{deltaT} \Delta t(\theta)=
{{2D}\over{\bar{c}_\gamma(\theta)}}-
{{2D}\over{\bar{c}_\gamma(\pi/2+\theta)}} \sim
{{2D}\over{c}}~{{\Delta \bar{c}_\theta } \over{c}}
\end{equation}

This relation was at the base of the original Michelson
interferometer but is also valid today when we assume Lorentz
transformations. In this case, in fact, the length D, in the $S'$
frame where the rod is at rest, is not depending on the orientation
(in the last relation, we are assuming light propagation in a medium
with a refractive index ${\cal N}=1 + \epsilon$, and $\epsilon\ll
1$). We thus get the fringe patterns ($\lambda$ being the wavelength
of light)
\begin{equation}
\label{newintro} {{\Delta \lambda(\theta)}\over{\lambda}} \sim
{{2D}\over{\lambda}} ~{{\Delta \bar{c}_\theta } \over{c}}
\end{equation}
which were measured in the old experiments.

\begin{figure}
\begin{center}
\includegraphics[width=10.5 cm]{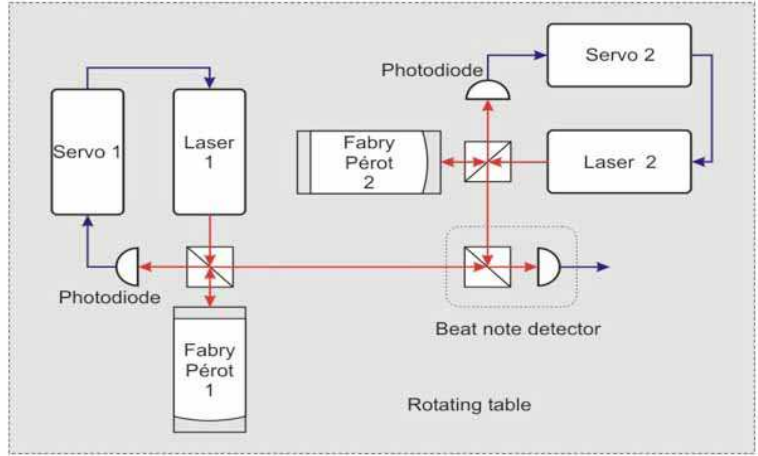}
\caption{The scheme of a modern ether-drift experiment. The
light frequencies are first stabilized by coupling the lasers to
Fabry-Perot optical resonators. The frequencies $\nu_1$ and $\nu_2$
of the signals from the resonators are then compared in the beat
note detector which provides the frequency shift $\Delta \nu=\nu_1
-\nu_2$. }
\label{Fig.apparatus}
\end{center}
\end{figure}

Instead, nowadays, an angular dependence of $\bar{c}_\gamma(\theta)$
is extracted from the frequency shift $\Delta\nu(\theta)$ of two
optical resonators, see Fig.\ref{Fig.apparatus}. The particular type
of laser-cavity coupling used in the experiments is known in the
literature as the Pound-Drever-Hall system \cite{pound,PDH}. The
details of this technique go beyond our scopes. However, the main
ideas are simple and beautifully explained in Black's tutorial
article \cite{black}. A laser beam is sent into a Fabry-Perot cavity
which acts as a filter. Then, a part of the output of the cavity is
fed back to the laser to suppress its frequency fluctuations. This
method provide a very narrow bandwidth and has been crucial for the
precision measurements we are going to describe.

The frequency of the resonators is proportional to
$\bar{c}_\gamma(\theta)$ through an integer number $m$, fixing the
cavity mode, and the cavity length $L$ measured in the laboratory
$S'$ frame
\begin{equation}
\label{nutheta0}
  \nu (\theta)= {{ \bar{c}_\gamma(\theta) m}\over{2L}}
\end{equation}
Again, by assuming Lorentz transformations, the length of the
cavity, in its rest frame $S'$, does not depend on the orientation
in space, so that
\begin{equation}
\label{bbasic2}
 {{\Delta \nu(\theta) }\over{\nu_0}}  \sim
      {{\Delta \bar{c}_\theta } \over{c}}
\end{equation}
$\nu_0$ being the reference frequency of the resonators. These
relations have always been assumed in the interpretation of the
experiments.

\section{A modern version of Maxwell calculation}

For a quantitative analysis, let us consider a medium of refractive
index ${\cal N}= 1+ \epsilon$, with $0 \leq \epsilon \ll 1$. This
medium fills an optical cavity at rest in the laboratory $S'$ frame
in motion with velocity $v$ with respect to $\Sigma$. If we assume
a) that $\bar{c}_\gamma(\theta)$ is isotropic when $S'\equiv \Sigma$
and b) that Lorentz transformations are valid, then any anisotropy
in $S'$ should vanish identically either for $v = 0$ or for the
ideal vacuum limit, i.e. when the velocity of light tends to the
basic parameter $c$ of Lorentz transformations \footnote{However a
null result in an ideal vacuum can also be deduced \cite{guerra2005}
without assuming Lorentz transformations, but only from simple
assumptions on the choice of the admissible clocks.}. Therefore, one
can perform an expansion in powers of the two small parameters
$\epsilon$ and $\beta=v/c$. Since, by its definition,
$\bar{c}_\gamma(\theta)$ is invariant under the replacement $\beta
\to -\beta$ and, at fixed $\beta$, is invariant when replacing
$\theta \to \pi +\theta$, the lowest non-trivial angular dependence
is found to ${\cal O}(\epsilon\beta^2)$ and can be expressed in the
general form \cite{epjc,dedicated,foop}
\begin{eqnarray}
\label{legendre}
       \bar{c}_\gamma(\theta) \sim {{c}\over{ {\cal N} }} \left[1- \epsilon~\beta^2
\sum^\infty_{n=0}\zeta_{2n}P_{2n}(\cos\theta)
  \right]
\end{eqnarray}
In the above relation, the invariance under $\theta \to \pi +\theta$
is achieved by expanding in even-order Legendre polynomials with
arbitrary coefficients $\zeta_{2n}={\cal O}(1)$. These coefficients
vanish identically in Einstein's special relativity, with no
preferred system, but should not vanish {\it a priori} in a
``Lorentzian'' formulation.

If we retain the first few $\zeta$'s as free parameters,
Eq.(\ref{legendre}) could already be useful for fitting experimental
data. In any case, independently of their numerical values, one
should appreciate the substantial difference introduced with respect
to the classical prediction. As an example, by assuming for
simplicity $v=$ 300 km/s, $\epsilon \sim 2.8 \cdot 10^{-4}$ as for
air at room temperature and atmospheric pressure, we find a
difference \BE \label{huge}\frac {\Delta\bar{c}_\theta(0)}{c} \sim
2.8 \cdot 10^{-10} \left[{{3}\over{ 2 }} \zeta_2 + {{5}\over{ 8 }}
\zeta_4 + ...\right] \EE This would be about three orders of
magnitude smaller than the classical estimate $\beta^2=10^{-6}$,
expected from Maxwell's calculation \cite{maxwell}, for the same
$v=$ 300 km/s. However, depending on the actual $\zeta$'s,
Eq.(\ref{huge}) would also be about 10$\div$ 20 times smaller than
the old standard value for the much lower orbital velocity $v_{\rm
orb} =$30 km/s \BE \frac {\Delta\bar{c}_\theta(0)}{c}\Big|_{\rm
class}=\frac {v^2_{\rm orb}}{c^2} = 10^{-8} \EE For experiments in
gaseous helium at room temperature and atmospheric pressure, where
$\epsilon \sim 3.3 \cdot 10^{-5}$, the equivalent of Eq.(\ref{huge})
would even be $100\div200$ times smaller than this old standard. The
above elementary arguments suggest that the old ether-drift
experiments in gaseous media might have been overlooked. So far,
they have been considered as null results. But this may just depend
on comparing with the wrong classical formula.

Yet, the dependence on the unknown $\zeta$'s is unpleasant because
it prevents a straightforward comparison with the data. For this
reason, by other symmetry arguments \cite{plus,plus2,book}, we will
further sharpen our analysis with another derivation of the
$\epsilon \to 0$ limit. This additional derivation makes use of the
effective space-time metric $g^{\mu\nu}=g^{\mu\nu}({\cal N})$ which
should be replaced into the relation $g^{\mu\nu}p_\mu p_\nu=0$ to
describe light in a medium of refractive index ${\cal N}$, see e.g.
\cite{leonhardt}. At the quantum level, this metric was derived by
Jauch and Watson \cite{jauch} when quantizing the electromagnetic
field in a dielectric. They realized that the formalism introduces a
preferred reference system where the photon energy $E_\gamma$ does
not depend on the angle $\theta$ of light propagation. They observed
that this is ``usually taken as the system for which the medium is
at rest'', a conclusion which is obvious in special relativity where
there is no preferred system but less obvious in our case. We will
therefore adapt their results and consider the different limit where
the photon energy $E_\gamma$ is $\theta-$independent only when {\it
both} medium and observer are at rest in some frame $\Sigma$.

To see how this works, we will consider two identical optical
resonators, namely resonator 1, which is at rest in $\Sigma$, and
resonator 2, which is at rest in $S'$. We will also introduce
$\pi_\mu\equiv ( {{E_\pi}\over{c}},{\bf \pi}) $, to indicate the
light 4-momentum for $\Sigma$ in his cavity 1, and $p_\mu\equiv (
{{E_p}\over{c}},{\bf p})$, to indicate the analogous 4-momentum of
light for $S'$ in his cavity 2. Finally we will define by
$g^{\mu\nu}$ the space-time metric used by $S'$ in the relation
$g^{\mu\nu}p_\mu p_\nu=0$ and by
\begin{equation}
\label{metricsigma}\gamma^{\mu\nu}={\rm diag}({\cal N}^2,-1,-1,-1)
\end{equation}
the metric which $\Sigma$ adopts in the analogous relation
$\gamma^{\mu\nu}\pi_\mu\pi_\nu=0$ and which produces the isotropic
velocity $c_\gamma=E_\pi/|{\bf \pi}|={{c}\over{{\cal N}}}$.

We emphasize the peculiar view of special relativity where no
observable difference can exist between $\Sigma$ and $S'$. In our
perspective, instead, this physical equivalence is only assumed in
the ideal ${\cal N}=1$ vacuum limit. Indeed, as anticipated in the
Introduction, in the presence of matter, where light gets absorbed
and then re-emitted, the fraction of refracted light could keep
track of the particular motion of matter with respect to $\Sigma$
and produce a $\Delta \bar{c}_\theta \neq 0$.

By first considering the ${\cal N}=1$ limit, the frame-independence
of the velocity of light requires to impose
\begin{equation} \label{limitingintro} g^{\mu\nu}({\cal N}=1)=
\gamma^{\mu\nu}({\cal N}=1)=\eta^{\mu\nu}\end{equation} where
$\eta^{\mu\nu}$ is the Minkowski tensor. This standard equality
amounts to introduce a transformation matrix, say $A^{\mu}_{\nu}$,
such that, for ${\cal N}=1$
\begin{equation}
\label{vacuum} g^{\mu\nu}({\cal
N}=1)=A^{\mu}_{\rho}A^{\nu}_{\sigma}\gamma^{\rho\sigma}({\cal N}=1)=
A^{\mu}_{\rho}A^{\nu}_{\sigma}\eta^{\rho\sigma}=\eta^{\mu\nu}
\end{equation}
This relation is strictly valid for ${\cal N}=1$. However, by
continuity, one is driven to conclude that an analogous relation
between $g^{\mu\nu}$ and $\gamma^{\mu\nu}$ should also hold in the
$\epsilon \to 0$ limit. The only subtlety is that relation
(\ref{vacuum}) does not fix uniquely $A^{\mu}_{\nu}$. In fact, it is
fulfilled either by choosing the identity matrix, i.e.
$A^{\mu}_{\nu}=\delta^{\mu}_{\nu}$, or by choosing a Lorentz
transformation, i.e. $A^{\mu}_{\nu}=\Lambda^{\mu}_{\nu}$. It thus
follows that $A^{\mu}_{\nu}$ is a two-valued function when ${\cal N}
\to 1$.

This gives two possible solutions for the metric in $S'$. In fact,
when $A^{\mu}_{\nu}$ is the identity matrix, we find
\begin{equation}\left[g^{\mu\nu}({\cal N})\right]_1=
\delta^{\mu}_{\rho}\delta^{\nu}_{\sigma}\gamma^{\rho\sigma}=
\gamma^{\mu\nu}\sim \eta^{\mu\nu} + 2\epsilon \delta^\mu_0
\delta^\nu_0\end{equation} while, when $A^{\mu}_{\nu}$ is a Lorentz
transformation, we obtain
\begin{equation} \label{2intro} \left[g^{\mu\nu}({\cal
N})\right]_2= \Lambda^{\mu}_{\rho}
\Lambda^{\nu}_{\sigma}\gamma^{\rho\sigma}\sim \eta^{\mu\nu} +
2\epsilon v^\mu v^\nu
\end{equation} where $v_\mu$ is the
$S'$ 4-velocity, $v_\mu\equiv(v_0,{\bf v}/c)$ with $v_\mu v^\mu=1$.
As a consequence, the equality $\left[g^{\mu\nu}({\cal
N})\right]_1=\left[g^{\mu\nu}({\cal N})\right]_2$ can only hold for
$v^\mu=\delta^\mu_0$, i.e. for ${\bf v}=0$ when $S'\equiv \Sigma$.

Notice that by choosing the first solution $\left[g^{\mu\nu}({\cal
N})\right]_1$, which is implicitly assumed in special relativity to
preserve isotropy in all reference frames also for ${\cal N} \neq
1$, we are considering a transformation matrix $A^{\mu}_{\nu}$ which
is discontinuous for any $\epsilon \neq 0$. In fact, it is the
non-trivial peculiarity of Lorentz transformations to enforce
Eq.(\ref{vacuum}) for $A^{\mu}_{\nu}=\Lambda^{\mu}_{\nu}$ so that
$\Lambda^{\mu \sigma}\Lambda^{\nu}_{\sigma}=\eta^{\mu\nu}$ and the
Minkowski metric, if valid in one frame, will then apply to all
equivalent frames.

On the other hand, if one inserts  $\left[g^{\mu\nu}({\cal
N})\right]_2$ in the relation $p_\mu p_\nu g^{\mu\nu}=0$, the photon
energy $E(| {\bf{p}}| , \theta)$ will now depend on the direction of
propagation. This gives the one-way velocity $c_\gamma(\theta)={{E(|
{\bf{p}}| , \theta)}\over{|{\bf{p}}|}}$ which, to ${\cal
O}(\epsilon)$ and ${\cal O}(\beta^2)$, is
\BE \label{oneway0}
       c_\gamma(\theta) \sim {{c} \over{{\cal N}}}~\left[
       1- 2\epsilon \beta \cos\theta -
       \epsilon \beta^2(1+\cos^2\theta)\right]
\EE
with a two-way combination
\begin{equation}
\label{twoway0}
       \bar{c}_\gamma(\theta)=
       {{ 2  c_\gamma(\theta) c_\gamma(\pi + \theta) }\over{
       c_\gamma(\theta) + c_\gamma(\pi + \theta) }} \sim
       {{c} \over{{\cal N} }}~\left[1-\epsilon\beta^2 (1 +
       \cos^2\theta) \right]
\end{equation} This final form, corresponding to setting in Eq.(\ref{legendre})
$\zeta_0=4/3$, $\zeta_{2}= 2/3$ and all $\zeta_{2n}=0$ for $n
> 1$, could be considered the modern version of Maxwell's calculation
\cite{maxwell} and  will be adopted in the analysis of experiments
near the $\epsilon = 0$ limit, as for gaseous systems.

For sake of clarity, let us  return to the definition of the gas
refractive index ${\cal N}$ in Eq.(\ref{metricsigma}). How, is this
quantity related to the experimental value ${\cal N}_{\rm exp}$
obtained from the two-way velocity measured in the earth laboratory?
This can be easily understood by first introducing an
angle-dependent $\bar{\cal N}(\theta)$ through
$\bar{c}_\gamma(\theta) \equiv c/\bar{\cal N}(\theta)$ with \BE
\label{nbartheta} \bar{\cal N}(\theta) \sim {\cal N} ~\left[1+({\cal
N}-1)\beta^2 (1 +
       \cos^2\theta)\right] \EE and then defining the
isotropic experimental value after an angular averaging, namely
\begin{equation} \label{nexp} {{c}\over{ {\cal N}_{\rm exp} }}\equiv
\langle {{c}\over{\bar{\cal N}(\theta)}} \rangle_\theta= {{c}\over{
{\cal N}  }} ~\left[1-{{3}\over{2}} ({\cal N} -1)\beta^2\right]
\end{equation}
One could therefore obtain the unknown value ${\cal N} \equiv {\cal
N}(\Sigma)$ (as if the cavity with the gas were at rest in
$\Sigma$), from the experimental ${\cal N}_{\rm exp}\equiv{\cal
N}(earth)$ and $v$. As an example, the most precise determinations
for air are at a level $10^{-7}$, say ${\cal N}_{\rm
exp}=1.0002924..$ for $\lambda=$ 589 nm, 0 $^o$C and atmospheric
pressure. Therefore, for $ v \sim $ 370 km/s, the difference between
${\cal N}(\Sigma)$ and ${\cal N}(earth)$ is smaller than $10^{-9}$
and can be ignored. Analogous considerations apply to other gaseous
media (as N, CO2, helium,..) where the precision  in ${\cal N}_{\rm
exp}$ is, at best, at the level of a few $10^{-7}$. Finally,
whatever $v$, the relation ${\cal N}(\Sigma)={\cal N}_{\rm exp}$
becomes more and more accurate in the low-pressure limit where
$({\cal N}_{\rm exp}-1)\to 0$.

To conclude, from Eq.(\ref{twoway0}) the fractional anisotropy is
found to be
\begin{equation} \label{bbasic2new} {{\Delta \bar{c}_\theta }
\over{c}}
={{\bar{c}_\gamma(\pi/2+\theta)-\bar{c}_\gamma(\theta)}\over{c}}\sim
     \epsilon~
       {{v^2 }\over{c^2}} \cos2(\theta-\theta_2) \end{equation}
and is suppressed by the small factor $2\epsilon$ with respect to
the classical estimate ${{\Delta \bar{c}_\theta } \over{c}} \sim
{{v^2 }\over{2c^2}}$. Here $v$ and $\theta_2$ indicate the magnitude
and the direction of the drift in the interferometer's plane and,
from Eq.(\ref{newintro}), one obtains the fringe pattern
\begin{equation} \label{newintro1} {{\Delta
\lambda(\theta)}\over{\lambda}}= {{2D}\over{\lambda}} ~{{\Delta
\bar{c}_\theta } \over{c}}\sim {{2D}\over{\lambda}}~
\epsilon~{{v^2}\over{c^2}}\cos 2(\theta -\theta_2)
\end{equation}   In this way, the dragging of light in the earth frame
is described as a pure 2nd-harmonic effect which is periodic in the
range $[0,\pi]$. This is the same as in the classical theory (see
e.g. \cite{kennedy}), with the exception of its amplitude
 \begin{equation} \label{a2new}
 A_2={{2 D}\over{\lambda}} ~\epsilon ~{{{v}^2}\over{c^2}}
\end{equation}
which is suppressed by the factor $2\epsilon$ relatively to the
classical amplitude $A^{\rm class}_2={{ D
}\over{\lambda}}{{{v}^2}\over{c^2}}$. This difference could then be
re-absorbed into an {\it observable} velocity
 \begin{equation} \label{aa2new}
 A_2={{ D}\over{\lambda}}{{{v}^2_{\rm obs}}\over{c^2}}
\end{equation}
which depends on the gas refractive index
\begin{equation}  \label{vobs} v^2_{\rm obs} \sim 2\epsilon v^2 \end{equation}
This $v_{\rm obs}$ is the very small velocity traditionally
extracted from the classical analysis of the experiments through the
relation
\begin{equation}  \label{vobs2} v_{\rm obs} \sim 30 ~{\rm km/s} ~\sqrt{\frac {A^{\rm EXP}_2}{A^{\rm class}_2}}\end{equation}
when one was still comparing with the standard classical prediction
$A^{\rm class}_2={{ D}\over{\lambda}} (\frac{30~{\rm km/s}}{c})^2$
for the orbital velocity.

However, before a more detailed comparison with experiments,
additional considerations are needed about the physical nature of
the ether-drift as an irregular phenomenon. Some general motivations
and a simple stochastic model will be illustrated in the following
section.

\section{Dragging of light as an irregular phenomenon}

Besides the magnitude of the signal, the other important aspect of
the experiments concerns the time dependence of the data. As
anticipated in the Introduction, it was always assumed that, for
short-time observations of a few days, where there are no sizeable
changes in the orbital velocity of the earth, a genuine physical
signal should reproduce the regular modulations induced by its
rotation. The data instead, for both classical and modern
experiments, have always shown a very irregular behavior with
statistical averages much smaller than the instantaneous values.
This was always a strong argument to interpret the data as
instrumental artifacts. However, in principle, a definite {\it
instantaneous} value $ {{\Delta\bar{c}_\theta(t)}\over{c}} \neq 0$
could also coexist with a vanishing statistical average.

This possibility was considered in refs.\cite{plus} $-$
\cite{physica}by assuming that the observed signal is determined by
a local velocity field, say $v_\mu(t)$, which does {\it not}
coincide with the projection of the global earth motion, say $\tilde
v_\mu (t)$, at the observation site. By comparing with the motion of
a body in a fluid, the equality $v_\mu(t)=\tilde v_\mu(t)$ amounts
to assume a form of regular, laminar flow where global and local
velocity fields coincide. Instead, in the case of a turbulent fluid
large-scale and small-scale flows would only be related {\it
indirectly}.

An intuitive motivation for this turbulent-fluid analogy derives
from comparing the vacuum to a fluid with vanishing viscosity. Then,
within the Navier-Stokes equation, a laminar flow is by no means
obvious due to the subtlety of the zero-viscosity (or infinite
Reynolds number) limit, see for instance the discussion given by
Feynman in Sect. 41.5, Vol.II of his Lectures \cite{feybook}. The
reason is that the velocity of such hypothetical fluid cannot be a
differentiable function \cite{onsager} and one should think,
instead, in terms of a continuous, nowhere differentiable velocity
field \cite{eyink}. This analogy leads to the idea of a signal with
a fundamental stochastic nature as when turbulence, at small scales,
becomes homogeneous and isotropic.

With this in mind, let us return to Eq.(\ref{bbasic2new}) and make
explicit the time dependence of the signal by re-writing as
\begin{equation} \label{basic2}
     {{\Delta \bar{c}_\theta(t) } \over{c}}
    \sim
 \epsilon {{v^2(t) }\over{c^2}}\cos 2(\theta
-\theta_2(t)) \end{equation}  where $v(t)$ and $\theta_2(t)$
indicate respectively the {\it instantaneous} magnitude and
direction of the drift in the plane of the interferometer. This can
also be re-written as
\begin{equation} \label{basic3} {{\Delta \bar{c}_\theta(t) } \over{c}}\sim
2{S}(t)\sin 2\theta +
      2{C}(t)\cos 2\theta \end{equation} with \begin{equation} \label{amplitude10}
       2C(t)= \epsilon~ {{v^2_x(t)- v^2_y(t)  }
       \over{c^2}}~~~~~~~2S(t)=\epsilon ~{{2v_x(t)v_y(t)  }\over{c^2}}
\end{equation}  and $v_x(t)=v(t)\cos\theta_2(t)$, $v_y(t)=v(t)\sin\theta_2(t)$

The standard analysis is based on a cosmic velocity of the earth
characterized by a magnitude $V$, a right ascension $\alpha$ and an
angular declination $\gamma$. These parameters can be considered
constant for short-time observations of a few days so that, with the
traditional identifications $ v(t)\equiv \tilde v(t)$ and
$\theta_2(t)\equiv\tilde\theta_2(t)$, the only time dependence
should be due to the earth rotation. Here $ \tilde v(t)$ and
$\tilde\theta_2(t)$ derive from the simple application of spherical
trigonometry \cite{nassau}
\begin{equation} \label{nassau1}
       \cos z(t)= \sin\gamma\sin \phi + \cos\gamma
       \cos\phi \cos(t'-\alpha)
\end{equation} \begin{equation} \label{projection}
       \tilde{v}(t) =V \sin z(t)
\end{equation} \begin{equation} \label{nassau2}
    \tilde{v}_x(t) = \tilde{v}(t)\cos\tilde\theta_2(t)= V\left[ \sin\gamma\cos \phi -\cos\gamma
       \sin\phi \cos(t'-\alpha)\right]
\end{equation} \begin{equation} \label{nassau3}
      \tilde{v}_y(t)= \tilde{v}(t)\sin\tilde\theta_2(t)= V\cos\gamma\sin(t'-\alpha) \end{equation}
In the above relations, $z=z(t)$ is the zenithal distance of
${\bf{V}}$, $\phi$ is the latitude of the laboratory,
$t'=\omega_{\rm sid}t$ is the sidereal time of the observation in
degrees ($\omega_{\rm sid}\sim {{2\pi}\over{23^{h}56'}}$) and the
angle $\tilde\theta_2(t)$ is counted conventionally from North
through East so that North is $\tilde\theta_2=0$ and East is
$\tilde\theta_2=90^o$. With the identifications $ v(t)\equiv \tilde
v(t)$ and $\theta_2(t)\equiv\tilde\theta_2(t)$ (or equivalently
$v_x(t)=\tilde{v}_x(t)$ and $v_y(t)=\tilde{v}_y(t)$) one thus
arrives to the simple Fourier decomposition
\begin{equation} \label{amorse1}
      S(t)\equiv \tilde{S}(t) =S_0+
      {S}_{s1}\sin t' +{S}_{c1} \cos t'
       + {S}_{s2}\sin(2t') +{S}_{c2} \cos(2t')
\end{equation}
\begin{eqnarray}
 \label{amorse2}
       C(t)\equiv  \tilde{C}(t)=
       {C}_0 + {C}_{s1}\sin t' +{C}_{c1} \cos t'
       + {C}_{s2}\sin(2 t') +{C}_{c2} \cos(2 t')
\end{eqnarray}
where the  $C_k$ and $S_k$ Fourier coefficients depend on the three
parameters $(V,\alpha,\gamma)$ and are given explicitly in
refs.\cite{plus,book}.

Instead, we will consider an alternative scenario where
$v_x(t)\neq\tilde{v}_x(t)$ and $v_y(t)\neq \tilde{v}_y(t)$. In
particular the local velocity components, $v_x(t)$ and $v_y(t)$,
will be assumed to be non-differentiable functions expressed in
terms of random Fourier series \cite{onsager,landau,fung}. The
simplest model corresponds to a turbulence which, at small scales,
appears homogeneous and isotropic. The analysis of the previous
section, can then be embodied in an effective space-time metric for
light propagation
\begin{equation} \label{random} g^{\mu\nu}(t) \sim \eta^{\mu\nu} + 2
\epsilon v^\mu(t) v^\nu(t) \end{equation} where $v_\mu(t)$ is a
random 4-velocity field which describes the drift and whose
boundaries depend on the smooth $\tilde{v}_\mu(t)$ determined by the
average motion of the earth. If this corresponds to the actual
physical situation, a genuine stochastic signal can easily become
consistent with average values $(C_k)^{\rm avg} = (S_k)^{\rm avg} =
0$ obtained by fitting the data with Eqs.(\ref{amorse1}) and
(\ref{amorse2}).

For homogeneous turbulence a series representation, suitable for
numerical simulations of a discrete signal, can be expressed in the
form
\begin{equation} \label{vx} v_x(t_k)= \sum^{\infty}_{n=1}\left[
       x_n(1)\cos \omega_n t_k + x_n(2)\sin \omega_n t_k \right] \end{equation}
\begin{equation} \label{vy} v_y(t_k)= \sum^{\infty}_{n=1}\left[
       y_n(1)\cos \omega_n t_k + y_n(2)\sin \omega_n t_k \right] \end{equation}
Here $\omega_n=2n\pi/T$ and T is the common period of all Fourier
components. Furthermore, $t_k= (k-1) \Delta t$, with $k=1, 2...$,
and $\Delta t$ is the sampling time. Finally, $x_n(i=1,2)$ and
$y_n(i=1,2)$ are random variables with the dimension of a velocity
and vanishing mean.

In our simulations, the value $T=T_{\rm day}$= 24 hours and a
sampling step $\Delta t=$ 1 second were adopted. However, the
results would remain unchanged by any rescaling $T\to s T$ and
$\Delta t\to s \Delta t$.

In general, we define $[-d_x(t),d_x(t)]$ the range for $x_n(i=1,2)$
and by $[-d_y(t),d_y(t)]$ the corresponding range for $y_n(i=1,2)$.
By assuming statistical isotropy we should impose $d_x(t)= d_y(t)$.
However, to see the difference, we will first consider the more
general case $d_x(t) \neq d_y(t)$. If we assume that $x_n(i=1,2)$
and $y_n(i=1,2)$ vary with uniform probability within their ranges
$[-d_x(t),d_x(t)]$ and $[-d_y(t),d_y(t)]$, the only non-vanishing
(quadratic) statistical averages are
\begin{equation} \label{quadratic} \langle x^2_n(i=1,2)\rangle_{\rm
stat}={{d^2_x(t) }\over{3 ~n^{2\eta}}}~~~~~~~~~~~\langle
y^2_n(i=1,2)\rangle_{\rm stat}={{d^2_y(t) }\over{3 ~n^{2\eta}}}
\end{equation} Here, the exponent $\eta$ ensures
finite statistical averages  $\langle v^2_x(t)\rangle_{\rm stat}$
and $\langle v^2_y(t)\rangle_{\rm stat}$ for an arbitrarily large
number of Fourier components. In our simulations, between the two
possible alternatives $\eta=5/6$ and $\eta=1$ of ref.\cite{fung}, we
have chosen $\eta=1$ that corresponds to the Lagrangian picture in
which the point where the fluid velocity is measured is a wandering
material point in the fluid.

In the end, the cosmic motion of the earth enters through the
identifications $d_x(t)= \tilde v_x(t)$ and $d_y(t)=\tilde v_y(t)$
as defined in Eqs.(\ref{nassau1})$-$(\ref{nassau3}) with $V=$ 370
km/s, $\alpha=168$ degrees, $\gamma= -$ 7 degrees, as fixed from our
motion within the CMB.

On the other hand, by assuming statistical isotropy, from the
relation
\begin{equation} \label{correct} \tilde{v}^2_x(t) +
\tilde{v}^2_y(t)=\tilde{v}^2(t)\end{equation} we obtain the
identification
\begin{equation} \label{isot} d_x(t)=d_y(t)={{ \tilde{v}(t)}\over{\sqrt{2} }} \end{equation}
For this isotropic model, from Eqs.(\ref{vx})$-$(\ref{isot}), we
find
\begin{eqnarray}
\label{vanishing} \langle v^2_x(t)\rangle_{\rm stat}=\langle
v^2_y(t)\rangle_{\rm stat}={{ \tilde v^2(t)}\over{2}}~{{1}\over{3}}
\sum^{\infty}_{n=1} {{1}\over{n^2}}= {{\tilde v^2(t)}\over{2}}~
{{\pi^2}\over{18}}\nonumber
\\ \langle v_x(t)v_y(t)\rangle_{\rm stat}=0
\end{eqnarray}
with statistical averages for the functions Eqs.(\ref{amplitude10})
\begin{equation} \label{vanishing2}\langle C(t)\rangle_ {\rm
stat}=0~~~~~~~~~~~~~~~~~~\langle S(t)\rangle_ {\rm stat}=0
\end{equation} which vanish at {\it any} time $t$. Therefore
this model describes a definite non-zero signal but, if this signal
were now fitted with Eqs.(\ref{amorse1}) and (\ref{amorse2}), it
would produce vanishing averages $(C_k)^{\rm avg}=0$, $(S_k)^{\rm
avg}=0$ for all Fourier coefficients. In other words, with such
physical signal, these statistical averages will become smaller and
smaller by simply increasing the number of observations.

\section{The classical experiments in gaseous media}

To understand how radical is the modification produced by
Eqs.(\ref{vanishing2}) in the analysis of the data, let us now
consider the traditional procedure adopted in the classical
experiments. One was measuring the fringe shifts at some given
sidereal time on consecutive days so that changes of the orbital
velocity were negligible. Then, see Eqs.(\ref{newintro1}) and
(\ref{basic3}), the measured shifts at the various angle $\theta$
were averaged
\begin{equation} \label{averagefringe}\langle{{\Delta
\lambda(\theta;t)}\over{\lambda}}\rangle_ {\rm stat}=
{{2D}\over{\lambda}} \left[2\sin 2\theta~\langle S(t)\rangle_ {\rm
stat} + 2\cos 2\theta~\langle C(t)\rangle_ {\rm stat} \right]
\end{equation} and finally these average values were compared with
models for the earth cosmic motion.

However if, following the arguments of the previous section, the
signal is so irregular that, by increasing the number of
measurements, $\langle C(t)\rangle_ {\rm stat} \to 0$ and $\langle
S(t)\rangle_ {\rm stat} \to 0$ the averages Eq.(\ref{averagefringe})
would have no meaning. In fact, these averages would be non
vanishing just because the statistics is finite. In particular, the
direction $\theta_2(t)$ of the drift (defined by the relation
$\tan2\theta_2(t)= S(t)/C(t)$) would vary randomly with no definite
limit.

\begin{table}
\caption{The 2nd-harmonic amplitudes for the six experimental
sessions of the Michelson-Morley experiment. The table is taken from
ref.\cite{plus}.}
\begin{center}
\begin{tabular}{cl}
\hline
SESSION       & ~~~~~~      $A^{\rm EXP}_2$   \\
\hline
July 8  (noon) & $0.010 \pm 0.005$  \\
July 9  (noon) & $0.015 \pm 0.005$   \\
July 11 (noon) & $0.025 \pm 0.005$    \\
July 8  (evening) & $0.014 \pm 0.005$  \\
July 9  (evening) &$0.011 \pm 0.005$   \\
July 12 (evening) & $0.024 \pm 0.005$  \\
\hline
\end{tabular}
\end{center}
\end{table}

Therefore, we should concentrate the analysis on the 2nd-harmonic
amplitudes \begin{equation} \label{AA} A_2(t)={{2D}\over{\lambda}}~
2\sqrt{S^2(t) + C^2(t)} \sim {{2 D }\over{\lambda}}~ \epsilon {{
v^2_x(t)+v^2_y(t)} \over{c^2 }}
\end{equation}
which are positive-definite and remain non-zero under the averaging
procedure. Moreover, these are rotational-invariant quantities and
their statistical properties would remain unchanged in the isotropic
model Eq.(\ref{isot}) or with the alternative choice $d_x(t)\equiv
\tilde v_x(t)$ and $d_y(t)\equiv \tilde v_y(t)$. In this way, in a
smooth deterministic model and using Eq.(\ref{projection}), we
obtain \BE \label{smoothfinal} \tilde A_2(t) \sim {{2 D
}\over{\lambda}} \cdot \epsilon{{ \tilde{v}^2(t) } \over{c^2 }} \sim
{{2 D }\over{\lambda}} \cdot \epsilon{{V^2  } \over{c^2 }} \cdot
\sin^2 z(t)\EE while, with a full statistical average, from
Eq.(\ref{vanishing}) \BE \label{amplitude10001final} \langle
A_2(t)\rangle_{\rm stat} \sim {{\pi^2 } \over{18 }}\cdot \tilde
A_2(t) \EE By comparing these two expressions, it is evident that,
from the same data, one would now get a velocity which is larger by
a factor $ \sqrt{ 18/\pi^2 }\sim$ 1.35. Also, from
Eq.(\ref{projection}), besides the average magnitude $\langle
\tilde{v}^2(t) \rangle_{\rm day}= V^2 \langle \sin^2 z(t)
\rangle_{\rm day}$, one could determine the angular parameters
$\alpha$ and $\gamma$ from the time modulations of the amplitude.

\begin{figure}
\begin{center}
\includegraphics[width=9.5 cm]{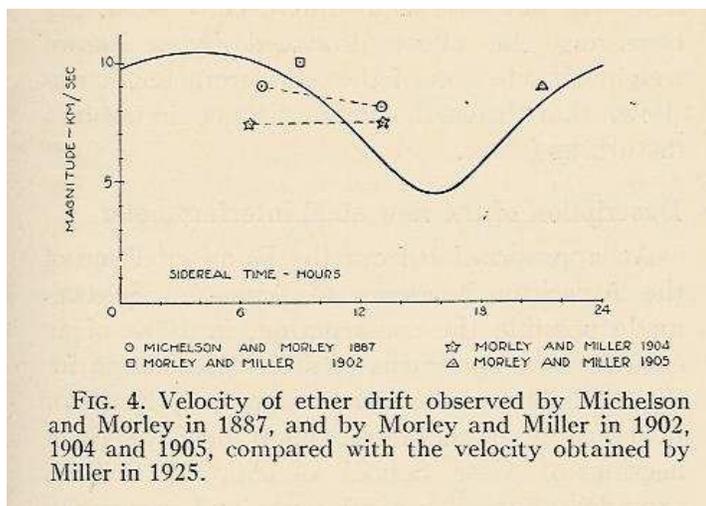}
\end{center}
\caption{The observable velocity measured in various experiments
as reported by Miller \cite{miller}.}
\label{miller}
\end{figure}

As an example, let us consider the 2nd-harmonic amplitudes for the
Michelson-Morley experiment, see Table 1. From these data, by
computing mean and variance, one finds $\langle A^{\rm EXP}_2
\rangle \sim 0.016 \pm 0.006$ so that by comparing with the
classical prediction $A^{\rm class}_2={{D}\over{\lambda}} {{(30 {\rm
km/s})^2 }\over{c^2}}\sim 0.20$ and using Eq.(\ref{vobs2}), we find
an observable velocity $ v_{\rm obs} \sim (8.4\pm 1.6) $ km/s in
good agreement with Miller's analysis, see Fig.\ref{miller}.
However, for air at atmospheric pressure where $\epsilon \sim
2.8\cdot 10^{-4}$, the true kinematical value would instead be $
\tilde v \sim (355 \pm 70)$ km/s from Eq.(\ref{smoothfinal}) or $
 \tilde v\sim (480 \pm 95)$ km/s from Eq.(\ref{amplitude10001final}).

Let us then consider Miller's very extensive observations. After the
re-analysis of his work by the Shankland team \cite{shankland},
there is now the average 2nd harmonic $\langle A^{\rm EXP}_2\rangle
=0.044 \pm 0.022$ for all epochs of the year (see Table III of
\cite{shankland}). By comparing this amplitude  with the classical
prediction for Miller's apparatus $A^{\rm
class}_2={{D}\over{\lambda}} {{(30 {\rm km/s})^2 }\over{c^2}}\sim
0.56$, we find $v_{\rm obs}\sim (8.4\pm 2.2)$ km/s. However, the
true kinematical velocity is instead $ \tilde v\sim (355 \pm 70)$
km/s from Eq.(\ref{smoothfinal}) or $ \tilde v\sim (480 \pm 95)$
km/s from Eq.(\ref{amplitude10001final}).

Note the agreement of two determinations obtained in very different
conditions (the basement of Cleveland laboratory or the top of Mount
Wilson). This shows that the traditional interpretation
\cite{joos2,shankland} of the residuals as temperature differences
in the optical paths is only acceptable provided these temperature
differences have a {\it non-local} origin. We will return to this
point in Section 6.

\begin{figure}
\begin{center}
\includegraphics[width=8.0 cm]{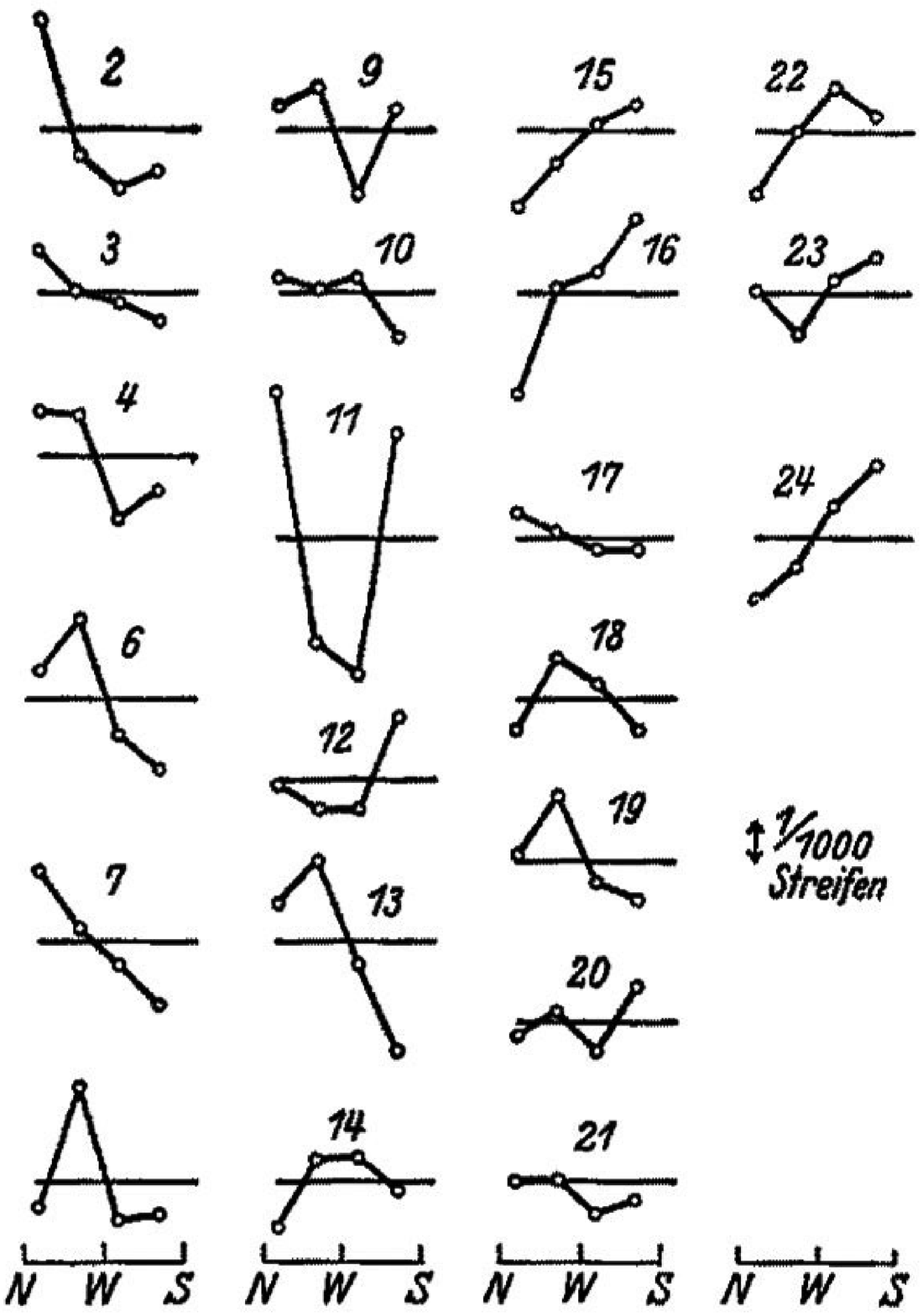}
\end{center}
\caption{The fringe shifts reported by Joos \cite{joos}. The yardstick corresponds to 1/1000 of a wavelength.}
\label{fringe-joos}
\end{figure}

There is no space for the details of all classical experiments. For
that, we address the reader to our book \cite{book} which also
contains many historical notes and references to previous works.
Here we will only limit ourselves to a brief description of Joos'
1930 experiment \cite{joos} in Jena (sensitivity of about 1/3000 of
a fringe) which is, by far, the most precise of the classical
repetitions of the Michelson-Morley experiment and is considered the
definitive disproof of Miller's claims of a non-zero effect
\footnote{Joos' optical system was enclosed in a hermetic housing
and, as reported by Miller \cite{miller,miller34}, it was
traditionally believed that his measurements were performed in a
partial vacuum. In his article, however, Joos is not clear on this
particular aspect. Only when describing his device for
electromagnetic fine movements of the mirrors, he refers to the
condition of an evacuated apparatus \cite{joos}. Instead, Swenson
\cite{swensonbook,loyd2} declares that Joos' fringe shifts were
finally recorded with optical paths placed in a helium bath.
Therefore, we have followed Swenson's explicit statements and
assumed the presence of gaseous helium at atmospheric pressure.}.

\begin{figure}
\begin{center}
\includegraphics[width=12.5 cm]{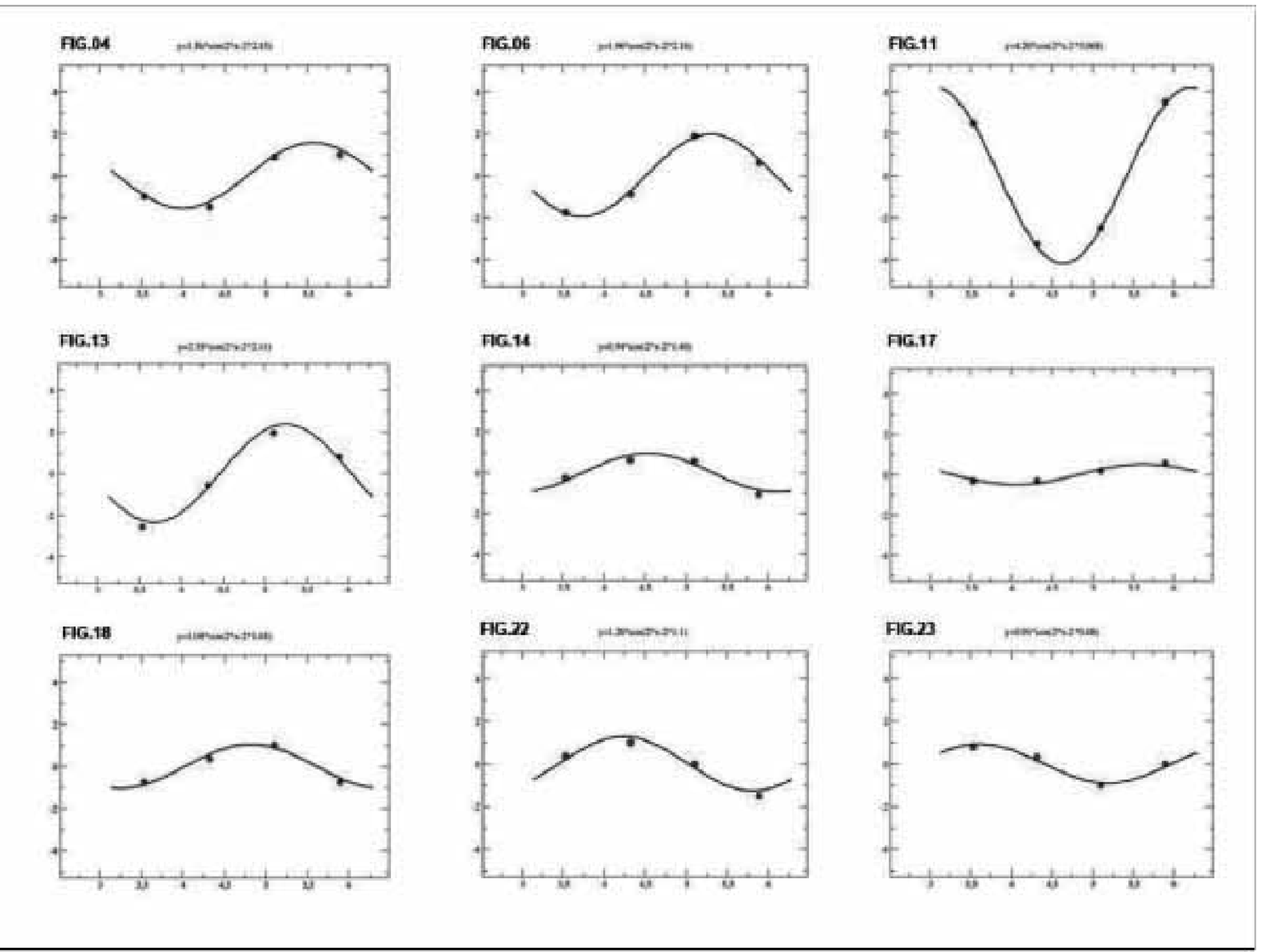}
\end{center}
\caption{ Some 2nd-harmonic fits to Joos' data. The figure is taken from
ref.\cite{book}.}
\label{collage}
\end{figure}

\begin{figure}
\begin{center}
\includegraphics[width=7.5 cm]{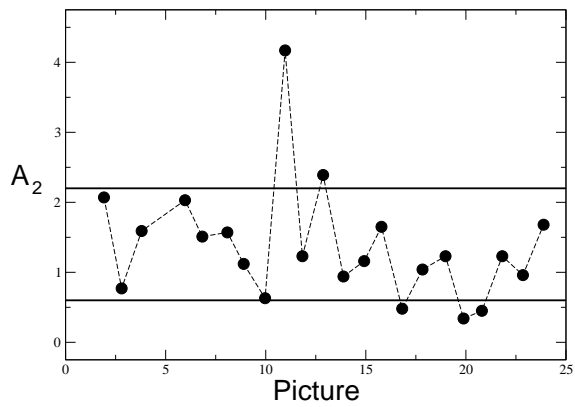}
\end{center}
\caption{Joos' 2nd-harmonic amplitudes, in units $10^{-3}$. The
vertical band between the two lines corresponds to the range $(1.4
\pm 0.8)\cdot10^{-3}$. The figure is taken from ref.\cite{plus}.}
\label{joosamplitudes}
\end{figure}

The data were taken at intervals of one hour during the sidereal day
and recorded photographically with an automatic procedure, see
Fig.\ref{fringe-joos}. From this picture, Joos adopted 1/1000 of a
wavelength as upper limit and deduced the bound $v_{\rm obs}
\lesssim 1.5$ km/s. To this end, he was comparing with the classical
expectation that, for his apparatus, a velocity of 30 km/s should
have produced a 2nd-harmonic amplitude of 0.375 wavelengths. Though,
since it is apparent that some fringe displacements were certainly
larger than 1/1000 of a wavelength, we have performed 2nd-harmonic
fits to Joos' data, see Fig.\ref{collage}. The resulting amplitudes
are reported in Fig.\ref{joosamplitudes}.

\begin{figure}
\begin{center}
\includegraphics[width=7.5 cm]{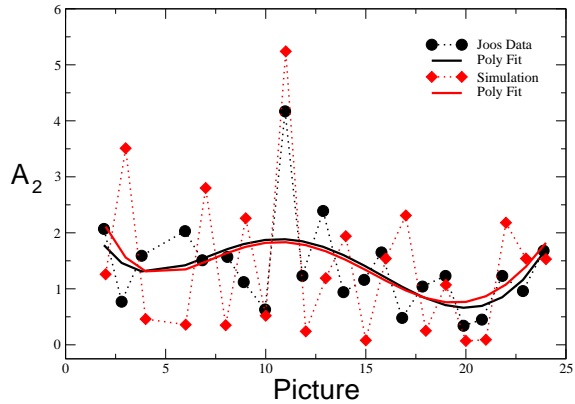}
\end{center}
\caption{Joos' 2nd-harmonic amplitudes, in units $10^{-3}$
(black dots), are compared with a single simulation (red diamonds)
at the same sidereal times of Joos' observations. Two 5th-order
polynomial fits to the two sets of values are also shown. The figure
is taken from ref.\cite{plus} .}
\label{joos-comparison}
\end{figure}

\begin{figure}
\begin{center}
\includegraphics[width=7.5 cm]{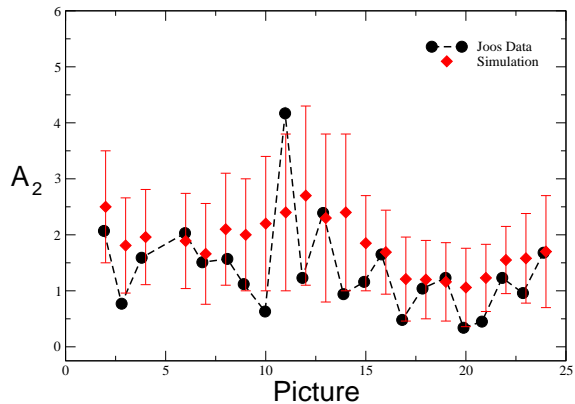}
\end{center}
\caption{Joos' 2nd-harmonic amplitudes in units $10^{-3}$ (black dots) are
now compared with a simulation where one averages ten measurements,
performed on 10 consecutive days, at the same sidereal times of
Joos' observations (red diamonds). The change of the averages
observed by varying the parameters of the simulation was summarized
into a central value and a symmetric error. The figure is taken from
ref.\cite{plus}.}
\label{joos-comparison-errors}
\end{figure}

We note that a 2nd-harmonic fit to the large fringe shifts in
picture 11 has a very good  chi-square, comparable and often better
than other observations with smaller values, see Fig.\ref{collage}.
Therefore, there is no reason to delete the observation n.11. Its
amplitude, however, is more than ten times larger than the
amplitudes from observations 20 and 21. This difference cannot be
understood in a smooth model of the drift where the projected
velocity squared at the observation site can at most differ by a
factor of two, as for the CMB motion at typical Central-Europe
latitude where $(\tilde v)_{\rm min} \sim 250$ km/s and $(\tilde
v)_{\rm max} \sim 370$ km/s. To understand these characteristic
fluctuations, we have thus performed various numerical simulations
of these amplitudes \cite{plus,book} in our stochastic model. To
this end, Eqs.(\ref{vx}) and (\ref{vy}) were replaced in
Eq.(\ref{AA}) and the random velocity components were bounded by the
kinematical parameters $(V,\alpha,\gamma)_{\rm CMB}$ as explained in
Sect.4. Two simulations are shown in Figs.\ref{joos-comparison} and
\ref{joos-comparison-errors}.

We want to emphasize two aspects. First, Joos' average amplitude
$\langle A^{\rm EXP}_2\rangle= (1.4 \pm 0.8)\cdot 10^{-3}$ when
compared with the classical prediction for his interferometer
$A^{\rm class}_2={{D}\over{\lambda}} {{(30 {\rm km/s})^2
}\over{c^2}}\sim 0.375$ gives indeed an observable velocity $v_{\rm
obs}\sim (1.8\pm 0.5)$ km/s very close to the $1.5$ km/s value
quoted by Joos. But, when comparing with our prediction in the
stochastic model Eq.(\ref{amplitude10001final}) one would now find a
true kinematical velocity $\tilde v= 305^{+85}_{-100}$ km/s. Second,
when fitting with Eqs.(\ref{nassau1}) and (\ref{projection}) the
smooth black curve of the Joos data  in Fig.\ref{joos-comparison}
one finds \cite{plus} a right ascension $\alpha({\rm fit-Joos})=
(168 \pm 30)$ degrees and an angular declination $\gamma({\rm
fit-Joos})= (-13 \pm 14)$ degrees which are consistent with the
present values $\alpha({\rm CMB}) \sim$ 168 degrees and $\gamma({\rm
CMB}) \sim -$7 degrees. This confirms that, when studied at
different sidereal times, the measured amplitude can also provide
precious information on the angular parameters.

Finally, all experiments are compared with our stochastic model
Eq.(\ref{amplitude10001final}) in Table \ref{summary}. Notice the
substantial difference with the analogous summary Table I of
ref.\cite{shankland} where those authors were comparing with the
classical relation $A^{\rm class}_2={{D}\over{\lambda}} {{(30 {\rm
km/s})^2 }\over{c^2}}$ and emphasizing the much smaller magnitude of
the experimental data. Here, is just the opposite. In fact, our
theoretical estimates are often {\it smaller} than the experimental
results indicating, most likely, the presence of systematic effects
in the measurements. At the same time, however, by adopting
Eq.(\ref{amplitude10001final}), the experiments in air give $\tilde
v_{\rm air} \sim 418 \pm 62 $ km/s and the two experiments in
gaseous helium $\tilde v_{\rm helium} \sim 323 \pm 70 $ km/s, with a
global average $\langle\tilde v \rangle\sim 376 \pm 46 $ km/s which
agrees well with the 370 km/s from the direct CMB observations. Even
more, from the most precise Piccard-Stahel and Joos experiments we
find two determinations, $\tilde v= 360^{+85}_{-110} $ km/s and
$\tilde v= 305^{+85}_{-100} $ km/s respectively, whose average
$\langle \tilde v\rangle \sim 332^{+60}_{-80} $ km/s reproduces to
high accuracy the projection of the CMB velocity at a typical
Central-Europe latitude \footnote{In ref.\cite{book} a numerical
simulation of the Piccard-Stahel experiment \cite{piccard3} is
reported, for both the individual sets of 10 rotations of the
interferometer and the experimental sessions (12 sets, each set
consisting of 10 rotations). Our analysis confirms their idea that
the optical path was much shorter than the instruments in United
States but their measurements were more precise because spurious
disturbances were less important.}.

\begin{table}
\caption{The average 2nd-harmonic amplitudes of classical
ether-drift experiments. These were extracted from the original
papers by averaging the amplitudes of the individual observations
and assuming the direction of the local drift to be completely
random (i.e. no vector averaging of different sessions). These
experimental values are then compared with the full statistical
average Eq.(\ref{amplitude10001final}) for a projection of the
velocity 250 km/s $\leq{\tilde v}(t) \leq$ 370 km/s and
refractivities $\epsilon=2.8\cdot10^{-4}$ for air and
$\epsilon=3.3\cdot10^{-5}$ for gaseous helium. The experimental
value for the Morley-Miller experiment is taken from the observed
velocities reported in Miller's Figure 4, here our Fig.\ref{miller}.
The experimental value for the Michelson-Pease-Pearson experiment
refers to the only known session for which the fringe shifts are
reported explicitly \cite{pease} and where the optical path was
still fifty-five feet. The symbol $\pm ....$ means that the
experimental uncertainty cannot be determined from the available
informations.}
\begin{tabular}{cllll}
\hline Experiment &gas
&~~~~$A^{\rm EXP}_2$ &~~~ ${{2D}\over{\lambda}}$~~~~& ~~~ $\langle A_2(t)\rangle_{\rm stat} $   \\
\hline
Michelson(1881)               & air     &$ (7.8 \pm....)\cdot10^{-3}$     &~~~$4\cdot 10^6  $   &$(0.7 \pm 0.2)\cdot 10^{-3}$  \\
Michelson-Morley(1887)   & air & $(1.6 \pm 0.6)\cdot 10^{-2 }$&~~~$4\cdot 10^7$ & $(0.7 \pm 0.2)\cdot 10^{-2}$ \\
Morley-Miller(1902-1905)   & air & $(4.0 \pm 2.0)\cdot 10^{-2 }$&~~~$1.12\cdot 10^8$ &$ (2.0 \pm 0.7) \cdot10^{-2}$\\
Miller(1921-1926)  & air& $(4.4 \pm 2.2)\cdot 10^{-2 }$ & ~~  $1.12\cdot 10^8$ &$(2.0 \pm 0.7) \cdot10^{-2} $ \\
Tomaschek (1924) & air & $(1.0\pm 0.6)\cdot 10^{-2 }  $ &~~~$3\cdot 10^7$& $ (0.5 \pm 0.2) \cdot10^{-2} $\\
Kennedy(1926)  & helium & ~~~$<0.002$&~~~$7 \cdot 10^6$&$ (1.4 \pm 0.5)\cdot10^{-4}  $\\
Illingworth(1927) & helium & $ (2.2 \pm 1.7)\cdot 10^{-4}  $  &~~~$7 \cdot 10^6$ &$ (1.4 \pm 0.5)\cdot10^{-4}$ \\
Piccard-Stahel(1928)          &air & $(2.8 \pm 1.5)\cdot10^{-3}$  &~~~$1.28 \cdot 10^7$& $(2.2 \pm 0.8)\cdot10^{-3}$\\
Mich.-Pease-Pearson(1929) & air& $(0.6 \pm....)\cdot10^{-2}$  &~~~$5.8  \cdot 10^7$& $(1.0 \pm 0.4)\cdot10^{-2}$\\
Joos(1930)  &helium&$(1.4 \pm 0.8)\cdot 10^{-3 }   $  & ~~ $7.5 \cdot 10^7$&$(1.5 \pm 0.6)\cdot10^{-3}$\\
\hline
\end{tabular}
\label{summary}
\end{table}

These non-trivial checks confirm the overall consistency of our
picture with the classical experiments and should induce to perform
new dedicated experiments where the optical resonators which are
coupled to the lasers (see Fig.\ref{Fig.apparatus}) are filled by
gaseous media. In this case, from Eq.(\ref{bbasic2}), one should
compare the data with the prediction
\begin{equation} \label{bbbasic2}
 {{\Delta \nu(\theta) }\over{\nu_0}}  =
      {{\Delta \bar{c}_\theta } \over{c}} \sim \epsilon~
       {{v^2 }\over{c^2}} \cos2(\theta-\theta_2) \end{equation}
However, precise measurements of the frequency shift in the gas mode
are not so simple \cite{stone}. For this reason, it is unclear if
there will be a definite improvement with respect to the classical
experiments, in particular with respect to Piccard-Stahel and Joos.

At present, a rough check of Eq.(\ref{bbbasic2}) can however be
obtained from the variations of the signal observed in the only
modern experiment that was performed in these conditions: the 1963
experiment by Jaseja et.al \cite{jaseja} with He-Ne lasers.
Actually, at that time, optical resonators were not yet used and
thus they were comparing directly the frequencies of two orthogonal
He-Ne lasers under 90 degrees rotations of the apparatus. But the
light from the lasers was emerging from a He-Ne gas mixture and thus
the laser frequencies were providing a measure of the two-way
velocity of light in that environment. As a matter of fact, for a
laser frequency $\nu_0\sim 2.6 \cdot 10^{14}$ Hz, after subtracting
a large systematic effect of about 270 kHz due to magnetostriction,
the residual variations of a few kHz are roughly consistent with the
refractive index ${\cal N}_{\rm He-Ne}\sim 1.00004$ and the typical
change of the cosmic velocity of the earth for the latitude of
Boston. For more details, see the discussion given in
\cite{epl,book}.
\section{Experiments in gases vs. vacuum and solid dielectrics}

The results in Table 2 support the idea of a tiny ${{\Delta
\bar{c}_\theta } \over{c}}$ at the level $10^{-10}$ for the
experiments in air and $10^{-11}$ for those in gaseous helium.
Simple symmetry arguments suggest the relation ${{\Delta
\bar{c}_\theta } \over{c}} \sim ({\cal N}_{\rm gas} -1){{v^2
}\over{c^2}}$ so that, from the data, we find the typical velocity
$v\sim $ 300 km/s expected from our motion within the CMB. But, yet,
one could ask: apart from symmetry arguments, how is the earth
motion producing this small observed anisotropy in the gaseous
systems?

As a possible hint, we recall that Eq.(\ref{legendre}) was
originally deduced in \cite{dedicated} as the most general angular
dependence of the refractive index in the presence of convective
currents of the gas molecules generated by an earth velocity $v$.
This idea of convection, with respect to the container of the gas at
rest in the laboratory, leads to reconsider the traditional
explanation of the small residuals in terms of tiny temperature
differences of a millikelvin or so \cite{joos2,shankland}. The
interesting aspect is that, besides helping our intuition, this
thermal interpretation will, in the end, be useful to analyze the
complementary region of solid dielectrics where the refractive index
${\cal N}$ is very different from unity.

In principle, as anticipated in the Introduction, with angular
differences $\Delta T^{\rm CMB}(\theta) = \pm$ 3.36 ~{\rm mK} of the
background radiation Eq.(\ref{CBR}), temperature differences of this
magnitude could reflect the collisions of the gas molecules, with
mean velocity of 370 km/s, with the CMB-photons. These collisions
could bring the gas out of equilibrium and induce a temperature
difference $\Delta T^{\rm gas}(\theta)$ in the optical paths. In
general, one expects $\Delta T^{\rm gas}(\theta) \leq \Delta T^{\rm
CMB}(\theta)$, the two extreme cases $\Delta T^{\rm gas}(\theta)=0$
and $\Delta T^{\rm gas}(\theta) = \Delta T^{\rm CMB}(\theta)$
corresponding respectively to the limits of vanishing interactions
or a very strong coupling of the two systems.

In view of the complexity of the calculation, we have not attempted
a full microscopic derivation of the effect but just limited
ourselves to a much simpler thermodynamic analysis
\cite{plus2,book}. This just assumes the existence of some $\Delta
T^{\rm gas}(\theta)$ to derive a corresponding difference in the
refractive index in the optical paths. Consistency with the idea of
a non-local effect will then require the same average $\langle
\Delta T^{\rm gas}(\theta)\rangle$ from different experiments.

This type of analysis starts from the Lorentz-Lorenz equation
\begin{eqnarray}\label{refra1}
        {{ {\cal N}^2 -1}\over{ {\cal N}^2 + 2  }} = A_R \rho + B_R
        \rho^2 ...
\end{eqnarray}
where $\rho$ is the molar density and $A_R= (4/3) \pi N_A \alpha$ is
the product of the Avogadro number $N_A$ and of the molecular
polarizability $\alpha$ (see e.g. \cite{stone}). The coefficient
$B_R$ takes into account two-body interactions which, for air and
helium at atmospheric pressure, can be ignored. For ${\cal N}\sim 1
$ we thus obtain the relation for the gas refractivity
\begin{eqnarray}\label{refra2}
       \epsilon= {\cal N} -1 \sim  {{ 3}\over{ 2 }}A_R \rho
\end{eqnarray}
In the ideal-gas approximation, the molar density at STP
(atmospheric pressure and $T=$ 273.15 K) has the value
\begin{eqnarray}\label{refra3}
        \rho(STP)=  {{ P}\over{ RT }} ={{ 101325}\over{ (8.314)
        (273.15)}}~{ \rm mol }\cdot { \rm m}^{-3 }\sim 4.46 \cdot 10^{-5 }~{ \rm mol }\cdot{ \rm cm}^{-3 }
\end{eqnarray}
As an example, for helium and a wavelength $\lambda=$ 633 nm, where
$\alpha\sim 0.52~ { \rm mol }^{-1 }\cdot{ \rm cm}^{3 }$
\cite{stone}, this gives $\epsilon\sim 3.5\cdot 10^{-5 }$. Thus, in
this simple approximation, where the temperature dependence of
$\epsilon$ is
\begin{eqnarray}\label{refra4}
-{{ \partial \epsilon }\over{ \partial T }} \sim {{ 3}\over{ 2 }}
A_R {{ P}\over{ RT^2 }} \sim {{ \epsilon}(T) \over{ T }}
\end{eqnarray}
and from the relation $\bar{c}_\gamma(\theta)
\equiv{{c}\over{\bar{\cal N}(\theta)}}$, a difference $\Delta T^{\rm
gas}(\theta)$ is seen to induce a typical angular difference \BE
\label{typical}{{|\Delta \bar{c}_\theta| } \over{c}} \sim ~
|\bar{\cal N}(\theta) -\bar{\cal N}(\pi/2+ \theta)| \sim {{~
\epsilon(T) |\Delta T^{\rm gas}(\theta)|} \over{ T }}\EE which
should be visible in the fringe shifts with a 2nd-harmonic amplitude
\begin{equation}
\label{newintroair} A^{ \rm exp}_2 \sim {{2D}\over{\lambda}}~
{{\epsilon(T) |\Delta T^{\rm gas} (\theta)|}\over{T}}
\end{equation}
For an average room temperature $T \sim 288\div 293$ K of the
experiments, the values of $|\Delta T^{\rm gas}(\theta)|$ are
reported in Table 3 for those cases where one can determine a
meaningful experimental uncertainty.
\begin{table}
\caption {The average 2nd-harmonic amplitude observed in various
classical ether-drift experiments and the resulting temperature
difference (in mK) from  Eq.(\ref{newintroair}).}
\begin{tabular}{cllll}
\hline Experiment &gas
&~~~~$A^{\rm EXP}_2$ &~~~ ${{2D}\over{\lambda}}$&  $|\Delta T^{\rm gas}(\theta)|$\\
\hline
Michelson-Morley(1887)   & air & $(1.6 \pm 0.6)\cdot 10^{-2 }$&~~~$4\cdot 10^7$ &$0.40 \pm 0.15$\\
Miller(1925-1926)  & air& $ (4.4 \pm 2.2)\cdot 10^{-2 } $ & ~~ $ 1.12\cdot 10^8$&$0.39 \pm 0.20$ \\
Illingworth(1927) & helium & $ (2.2 \pm 1.7)\cdot 10^{-4}  $  &~~~$7 \cdot 10^6$& $0.29\pm 0.22$\\
Tomaschek (1924) & air & $(1.0\pm 0.6)\cdot 10^{-2 }  $ &~~~$3\cdot 10^7$& $0.33 \pm 0.20$\\
Piccard-Stahel(1928)&air &$(2.8 \pm 1.5)\cdot 10^{-3 }$  &  ~~~$ 1.28 \cdot 10^7$& $0.22 \pm 0.12$\\
Joos(1930)  &helium&$(1.4 \pm 0.8)\cdot 10^{-3 }   $  & ~~ $7.5 \cdot 10^7$&$0.17 \pm 0.10$\\
\hline
\end{tabular}
\end{table}
The very good chi-square, 2.4/(6-1)=0.48, shows that all
experiments, can become consistent with the same average value \BE
\label{avetgas} \langle \Delta T^{\rm exp}(\theta) \rangle = (0.26
\pm 0.06) ~ {\rm mK} \EE so that the residuals observed in the old
experiments could also be interpreted as thermal effects of {\it
non-local} origin \footnote{We have not produced a microscopic
derivation from $\Delta T^{\rm CMB}= \pm 3.36$ mK, but still the
concordance of different experiments in different laboratories
suggests that our $\Delta T^{\rm gas}= (0.2\div 0.3)$ mK has a
fundamental origin. Interestingly, after a century from those old
experiments, in room-temperature measurements, the 1 mK level is
still state of the art for the precision attainable in temperature
differences, see e.g. \cite{farkas,zhaoa,trusov}.}.

This previous analysis suggests two considerations. First, the old
estimates of about 1 mK by Kennedy, Shankland and Joos (see
\cite{joos2,shankland}) were slightly too large. Within our present
view, this may indicate that the interactions of the gas molecules
with the CMB photons are so weak that, on average, only less than
1/10 of $\Delta T^{\rm CMB}(\theta)$ is transferred to the gas in
the optical paths. Second, with the thermal mechanism discussed
above, one could replace in Eq.(\ref{nbartheta}) $\epsilon_{\rm
gas}=({\cal N}_{\rm gas} -1)\equiv \epsilon_{\rm thermal} +
\epsilon_{v}$ and re-write
\begin{eqnarray}\label{6introgas}
       {{ \bar{\cal N}_{\rm gas}(\theta)}\over { {\cal N}_{\rm gas}}} \sim 1+(\epsilon_{\rm thermal}+ \epsilon_{
       v})\beta^2
        (1 +
       \cos^2\theta)
\end{eqnarray}
where $\epsilon_{\rm thermal}\equiv ({\cal N}_{\rm gas}- {\cal N}_v
)$, $\epsilon_{ v}\equiv ({\cal N}_v-1)$. In this way, we have
introduced an extremely small quantity $\epsilon_{v}$ which, in
principle, could still account for a difference between the velocity
of light $c_\gamma\equiv c/{\cal N}_v$, as measured in vacuum on the
earth surface, and the ideal parameter $c$ of Lorentz
transformations.

To roughly estimate a possible non-zero $\epsilon_v$, let us first
recall that today the (isotropic) speed of light in vacuum is a
reference standard with zero error, namely $c_{\rm ref}=$ 299 792
458 m/s and that the last precision measurements, performed before
fixing this reference value, had an error of about 1 m/s at the
3-sigma level \cite{nist}. Therefore assuming $|c- c_{\rm
ref}|\lesssim$ 1 m/s, we would tentatively estimate $\epsilon_v
\lesssim 10^{-9}$. As such, at room temperature and atmospheric
pressure, where this $\epsilon_v$ is numerically irrelevant,
$\epsilon_{\rm thermal}$ is practically the same refractive index
considered so far, i.e. $\epsilon_{\rm air}\sim 2.8\cdot 10^{-4}$ or
$\epsilon_{\rm helium}\sim 3.3\cdot 10^{-5}$. Nevertheless
Eq.(\ref{6introgas}) is useful because, in the opposite limit of an
extremely high vacuum where now $\epsilon_{\rm thermal}= 0 $, for
$\epsilon_v \neq 0$, we would predict the angular dependence
\begin{eqnarray}\label{6introvacuum}
       {{ \bar{\cal N}_v(\theta)}\over { {\cal N}_v}} \sim 1+\epsilon_{
       v}\beta^2
        (1 +
       \cos^2\theta)
\end{eqnarray} and an anisotropy of the
two-way velocity of light in vacuum
\begin{eqnarray}\label{6introany}
{{\Delta\bar{c}_\theta}\over{c}}\Big|_{\rm vacuum}= \bar{\cal
N}_v(\theta)- \bar{\cal N}_v(\pi/2 +\theta)
       ~\sim~ \epsilon_{
       v}\beta^2 \cos2\theta
\end{eqnarray}
Even more interestingly, the thermal argument is also useful to
analyze experiments in solid dielectrics, as that originally
performed by Shamir and Fox \cite{fox} in 1969. They were aware that
the Michelson-Morley experiment did not yield a strictly zero
result: ``The non-zero result might have been real and due to the
fact that the experiment was performed in air and not in vacuum''
\cite{fox}. Therefore, within the traditional Lorentz-contraction
interpretation of the experiment, with a refractive index ${\cal N}$
substantially above unity, one might expect a large
${{|\Delta\bar{c}_\theta|}\over{c}}\sim ({\cal N}^2-1)\beta^2\sim
\beta^2\sim 10^{-6}$. This was the motivation for their experiment
in perspex (${\cal N}=1.5$). Since their measurements were orders of
magnitude smaller, they concluded that the experimental basis of
special relativity was strengthened.

However, with a thermal interpretation of the residuals in gaseous
media, the two different behaviors can coexist. In fact, as
anticipated in the Introduction, in a strongly bound system as a
solid a small temperature gradient of a fraction of millikelvin
would mainly dissipate by heat conduction without any particle
motion or light anisotropy in the rest frame of the apparatus. On
this basis, with a very precise experiment, a fundamental vacuum
anisotropy as in Eq.(\ref{6introany}) could also become visible in a
solid dielectric.

To see how this works, let us first observe that, as in the gas
case, for ${\cal N}_v\neq 1$ there will be a very tiny difference
between the refractive index defined relatively to the ideal vacuum
value $c$ and the refractive index relatively to the physical
isotropic vacuum value $c/{\cal N}_v$ measured on the earth surface.
The relative difference between these two definitions is
proportional to $\epsilon_v \lesssim 10^{-9} $ and, for all
practical purposes, can be ignored. More significantly, all
materials would now exhibit the same background vacuum anisotropy
proportional to $ \epsilon_v \beta^2$ in Eq.(\ref{6introany}).  To
this end, let us first replace the average isotropic value \BE
{{c}\over { {\cal N}_{\rm solid}}} \to {{c}\over { {\cal N}_v {\cal
N}_{\rm solid} }}\EE and then use Eq.(\ref{6introvacuum}) to replace
${\cal N}_v$ in the denominator with $\bar {\cal N}_v(\theta)$. This
is equivalent to define a $\theta-$dependent refractive index for
the solid dielectric
\begin{eqnarray}\label{6introsolid}
        {{  \bar{\cal N}_{\rm solid}(\theta)}\over { {\cal N}_{\rm solid}}} \sim  1+\epsilon_v \beta^2(1
        +
       \cos^2\theta)
\end{eqnarray}
so that
\begin{equation}
\label{refractivetheta1} \left[ {\bar c_\gamma (\theta)}
\right]_{\rm solid}={{c}\over{\bar{\cal N}_{\rm
solid}(\theta)}}={{c}\over { {\cal N}_{\rm solid}}} \left[ 1-
\epsilon_v \beta^2 (1 +
       \cos^2\theta)\right]
\end{equation}
with an anisotropy
\begin{equation}
\label{anysolid} {{ \left[\Delta\bar{c}_\theta\right]_{\rm solid}}
\over {\left[ c/ {\cal N}_{\rm solid}\right] }} \sim \epsilon_v
\beta^2 \cos2\theta
\end{equation}
In this way, a genuine vacuum effect as in Eq.(\ref{6introany}), if
there, could also be detected with a very precise experiment in a
solid dielectric. It is then important to understand the magnitude
$\epsilon_v \lesssim 10^{-9} $ suggested by the last precision
measurements of about thirty years ago \cite{nist}. Is it just
accidental or does it express a fundamental property of light on the
earth surface? In the latter case, with $\epsilon_v \sim 10^{-9}
\neq 0 $ in Eqs.(\ref{6introany}) and (\ref{anysolid}), a typical
$10^{-15}$ beat signal should then show up. Let us therefore compare
with present experiments, starting from those with vacuum optical
resonators.

\section{Modern experiments with optical resonators}

\subsection{Basic aspects of present experiments in vacuum}

As anticipated, the Pound-Drever-Hall system \cite{pound,PDH} shown
in Fig.\ref{Fig.apparatus} was crucial for precision tests of
relativity. The first application dates back to Brillet and Hall in
1979 \cite{brillet}. They were comparing the frequency of a CH$_4$
reference laser (fixed in the laboratory) with the frequency of a
cavity-stabilized He-Ne laser placed on a rotating table. Since the
stabilizing optical cavity was placed inside a vacuum envelope, the
measured shift $\Delta \nu(\theta)$ was giving a measure of the
anisotropy of the velocity of light in vacuum.

In the last forty years, substantial improvements have been
introduced in the experiments. However, the assumptions behind the
analysis of the data are basically unchanged and any physical signal
is assumed to depend deterministically on the velocity of the earth
with respect to some fixed preferred frame. As emphasized in the
previous chapters, the macroscopic motion of the earth (i.e. on a
cosmic scale) could instead affect the microscopic propagation of
light in an optical cavity in some complicated, indirect way and a
genuine signal could easily be misinterpreted as a spurious effect.
For this reason, first of all, we will try to understand the
magnitude of the {\it instantaneous} signal with vacuum cavities and
then compare the data with numerical simulations performed within
the same model adopted for the classical experiments.

\begin{figure}
\begin{center}
\includegraphics[width=7.5 cm]{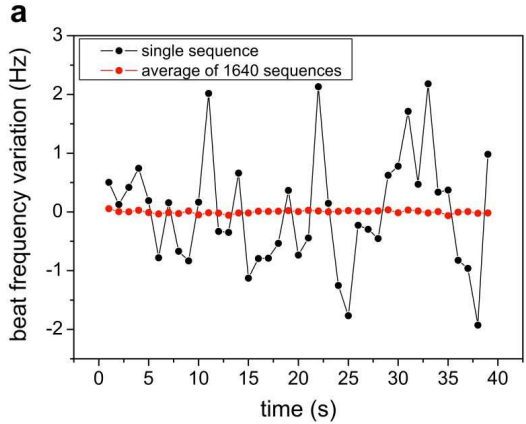}
\end{center}
\caption{The experimental frequency shift reported in Fig.9(a)
of ref.\cite{crossed} (courtesy Optics Communications). The black
dots give the instantaneous signal, the red dots give the signal
averaged over 1640 sequences. For a laser frequency $\nu_0=2.82\cdot
10^{14}$ Hz a $\Delta \nu=\pm 1$ Hz corresponds to a fractional
value $\Delta \nu/\nu_0$ of about $\pm 3.5 \cdot 10^{-15}$.}
\label{Figcrossed}
\end{figure}

To understand the magnitude of the signal we have compared with
Figure 9.a of ref.\cite{crossed} and Figure 4
ref.\cite{schiller2015}. These give the idea of a very irregular
$\Delta \nu$ with a typical magnitude in the range $\pm 1$ Hz, see
our Fig.\ref{Figcrossed}. For the adopted reference frequency
$\nu_0=2.8 \cdot 10^{14}$ Hz, this is the anticipated $10^{-15}$
fractional level. The same value is obtained from \cite{newberlin}.
Actually, in this other article the instantaneous signal is not
shown explicitly but it can be deduced from the typical variation
over a characteristic time of 1$\div$2 seconds. For an irregular
signal, in fact, this variation gives the magnitude of the signal
itself and its value is again $10^{-15}$.

After having obtained these first indications, we have tried to
understand the meaning of this irregular signal. Namely is it just
spurious noise (e.g. thermal noise \cite{numata}) or could it
represent a genuine signal? As a check, we have then compared with
other two experiments, ref.\cite{mueller2003} and
ref.\cite{cpt2013}, where the optical cavities were made of
different materials and were operating at a {\it cryogenic}
temperature. Again the same $10^{-15}$ level. Since it is extremely
unlike that spurious effects remain the same for experiments
operating in so different conditions, it is natural to explore the
possibility that such $10^{-15}$ signal admits a physical
interpretation.

Therefore, applying to the physical vacuum the same model used
successfully for the classical experiments, we will tentatively
express this observed fractional shift in terms of a cosmic earth
velocity and of a refractive index ${\cal N}_v$ as \BE
\label{anivacuum} \left|{{\Delta\nu(\theta)}\over
{\nu_0}}\right|_{\rm exp} =\left| {{\Delta\bar{c}_\theta}\over {c}}
\right|_{\rm exp} \sim ({\cal N}_v -1)~(v^2/c^2)~\sim {\cal O}
(10^{-15}) \EE For $v \sim$ 300 km/s, this supports our previous
idea of a tiny {\it refractivity} $\epsilon_v= ({\cal N}_v-1)\sim
10^{-9}$ for the physical vacuum established in an apparatus placed
on the earth surface.  Therefore, it is now the time to recall the
scenario of ref.\cite{gerg} which could indeed explain such result.

\subsection{A $10^{-9}$ refractivity for the vacuum on the earth
surface}

The idea of a non-zero vacuum refractivity may have different
motivations. The perspective of ref.\cite{gerg} was inspired by the
so called {\it emergent} gravity approach \cite{barcelo1}$-$
\cite{cosmo} where the introduction of a non-trivial metric field
$g_{\mu\nu}(x)$ is considered in analogy with the hydrodynamic limit
of many condensed-matter systems. This emergent interpretation is
made manifest in a parametric dependence of the metric on some
auxiliary, inducing-gravity fields $s_k(x)$, i.e. $g_{\mu\nu}(x)=
g_{\mu\nu}[s_k(x)]$. As in the pioneering Yilmaz derivation based on
the static Newtonian potential \cite{yilmaz,tupper}, Einstein
equations for the metric would then follow from the equations of
motion for the $s_k$'s in flat space, after introducing a suitable
stress tensor for these auxiliary fields. In this way, one could
(partially) fill the conceptual gap with classical General
Relativity.

An interesting consequence derives from the boundary condition
$g_{\mu\nu}[s_k=0]=\eta_{\mu\nu}$. In fact, if the $s_k$'s are
understood as {\it excitations} of the physical vacuum, which
therefore vanish identically in its equilibrium state, one could
easily understand \cite{volo} why the energy of the unperturbed
vacuum plays no role. This perspective of a non-gravitating vacuum
energy \cite{volo} provides, perhaps, the most intuitive solution of
the so called cosmological-constant problem usually mentioned in
connection with the quantum vacuum. In this sense, with this type of
approach, one is taking seriously Feynman's words: ``The first thing
we should understand is how to formulate gravity so that it doesn't
interact with the vacuum energy'' \cite{rule}.

Another interesting aspect of this approach is that, even without
knowing the underlying $s_k$'s, in the simplest case of a static
metric all dynamical effects are equivalent to two basic
ingredients: i) local modifications of the physical clocks and rods
and ii) local modifications of the velocity of light. Therefore,
with this interpretation of the observed curvature, one could try to
test the fundamental assumption of General Relativity that, in the
presence of gravity, the velocity of light in vacuum $c_\gamma$ is
still a universal constant, namely it remains the same, basic
parameter $c$ of Lorentz transformations. Notice that, here, we are
not considering the so called coordinate-dependent speed of light.
Rather, we are focused on the true, physical $c_\gamma$ as obtained
from experimental measurements in vacuum optical cavities placed on
the earth surface. Thus in principle, a precise measurement
establishing that $c_\gamma \neq c$ could give information on the
fundamental mechanisms at the base of the gravitational interaction.

For the various aspects of space-time measurements, a very clear
reference is Cook's article ``Physical time and physical space in
general relativity'' \cite{cook}. There, the appropriate units of
time and length, respectively $d\tau$ and $d l$, are defined to
ensure that all observers measure the same, universal speed of light
(``Einstein postulate''). For a static metric, these definitions are
$d\tau^2=g_{00} dt^2$ and $dl^2=g_{ij}dx^i dx^j$. Thus, in General
Relativity, the condition $ds^2=0$, which governs the propagation of
light, can be expressed formally as \BE ds^2= c^2d\tau^2-dl^2=0 \EE
and, by construction, gives always the same speed $dl/d\tau=c$.

But, if the physical units were instead $d\hat \tau$ and $d\hat l$
with, say, $d\tau = q~ d\hat\tau$ and $dl=p~ d\hat l$, the same
condition \BE ds^2= c^2q^2d \hat \tau^2-p^2 d\hat l^2=0 \EE would
now be interpreted differently as \BE c_\gamma={{ d\hat l}\over{
d\hat \tau }}=c ~{{q}\over{p}} \equiv {{c}\over { {\cal N}_v }} \EE
The possibility of different units is thus a simple motivation for a
vacuum refractive index ${\cal N}_v\neq 1$.

To fix the ideas, we will start from the unambiguous point of view
of special relativity: the right space-time units are those for
which the speed of light in the vacuum $c_\gamma$, when measured in
an inertial frame, coincides with the basic parameter $c$ of Lorentz
transformations. But inertial frames are just an idealization.
Therefore the physical realization is to assume standards of
distance and time which can change {\it locally} but such that the
identification $c_\gamma=c$ holds in the asymptotic condition which
is as close as possible to an inertial frame. This asymptotic
condition corresponds to measure $c_\gamma$ in a freely falling
frame \footnote{One should further restrict light propagation to a
small enough region that tidal effects of the external gravitational
potential $U_{\rm ext}(x)$ can be ignored.} and is crucial for an
operational definition of the otherwise {\it unknown} quantity $c$.

\begin{figure}
\begin{center}
\includegraphics[width=7.5 cm]{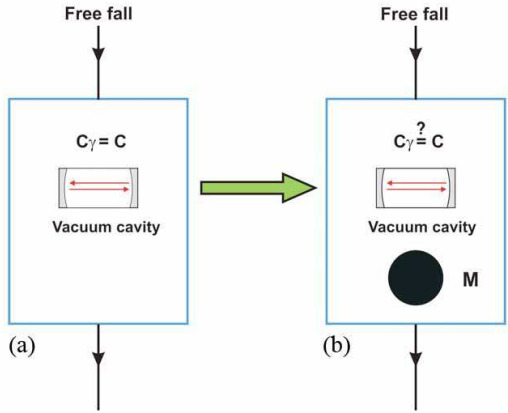}
\end{center}
\caption{An intuitive visualization of two physically distinct
situations. In case (b) a heavy mass $M$ is carried on board of a
freely-falling system. Differently from the ideal case (a), the mass
$M$ could introduce a vacuum refractivity so that now $c_\gamma \neq
c$.}
\label{freefall}
\end{figure}

With this premise, an observer $S'$ placed on the earth surface can
still describe light propagation in different ways. We address the
reader to ref.\cite{gerg}, where these aspects were originally
discussed, and to refs.\cite{plus2,book} for further refinements.
The whole idea, however, is simple and can be reduced to
Fig.\ref{freefall}. An observer $S'$ placed on the earth surface is
in free-fall with respect to all masses in the Universe but not with
respect to the gravitational field of the earth. Its effect can be
schematically represented by means of a heavy mass $M$ carried on
board of the elevator.

The two situations in panels {\bf (a)} and {\bf (b)} of
Fig.\ref{freefall} are physically distinct but in General Relativity
it is assumed that both observers will measure the same $c$ of
Lorentz transformations. A non-zero vacuum refractivity, for system
{\bf (b)}, can thus be expressed as
\begin{equation} \label{refractive}
\epsilon_v={\cal N}_v - 1 \sim {{z}\over{2}}~\left({{2|\delta
U|}\over{c^2}}\right)
\end{equation}
where $\delta U$ is the extra Newtonian potential produced by the
heavy mass $M$ at the experimental setup. In General Relativity one
assumes $z=0$ while the two non-zero values ($z=$ 1 or 2) account
for the two alternatives traditionally reported in the literature
for the effective refractive index in a gravitational potential (see
the discussion in refs.\cite{plus2,book} and in particular
Broekaert's footnote $^{3}$ \cite{broekaert}). In our case, by
introducing the Newton constant, the radius $R$ and the mass $M$ of
the earth, so that $\delta U={{G_N M}\over{R}}$, we find
\begin{equation} \label{refractive2}
\epsilon_v \sim {{z}\over{2}}~1.4\cdot 10^{-9} \end{equation} We
emphasize that, regardless of whether $z=$ 1 or 2, the velocity of
light in a vacuum cavity on the earth surface, panel {\bf (b)} in
our Fig.\ref{freefall}, could differ at the level $10^{-9}$ from
that ideal value $c$, operationally defined with the same apparatus
in a true freely-falling frame, panel {\bf (a)} in our
Fig.\ref{freefall}. As discussed at the end of Sect.6, such
$\epsilon_v\sim 10^{-9}$ was suggested by the last precise
measurements of the velocity of light and, by comparing with
Eq.(\ref{anivacuum}), could now provide a physical argument to
seriously consider the presently observed $10^{-15}$ fractional
frequency shift of two vacuum optical resonators. Let us therefore
give a closer look at the present experiments.

\subsection{A closer look at experiments and numerical simulation of the signal}

Most recent ether-drift experiments measure the frequency shift
$\Delta \nu$ of two {\it rotating} optical resonators. To this end,
let us re-write Eq.(\ref{basic2}) as
\begin{equation} \label{basic2new}
    {{\Delta \nu (t)}\over{\nu_0}} = {{\Delta \bar{c}_\theta(t) } \over{c}}
    \sim
 \epsilon {{v^2(t) }\over{c^2}}\cos 2(\omega_{\rm rot}t
-\theta_2(t)) \end{equation} where $\omega_{\rm rot}$ is the
rotation frequency of the apparatus. Therefore one finds
\begin{equation} \label{basic3new} {{\Delta \nu (t)}\over{\nu_0}}
\sim 2{S}(t)\sin 2\omega_{\rm rot}t +
      2{C}(t)\cos 2\omega_{\rm rot}t \end{equation} with $C(t)$ and $S(t)$
given in Eqs.(\ref{amplitude10}) for $\epsilon=\epsilon_v$
\begin{equation} \label{amplitude101}
       2C(t)= \epsilon_v~ {{v^2_x(t)- v^2_y(t)  }
       \over{c^2}}~~~~~~~2S(t)=\epsilon_v ~{{2v_x(t)v_y(t)  }\over{c^2}}
\end{equation}  and $v_x(t)=v(t)\cos\theta_2(t)$,
$v_y(t)=v(t)\sin\theta_2(t)$. For a non rotating apparatus, as in
Fig.\ref{Figcrossed}, the fractional frequency shift is thus simply
$2C(t)$.

\begin{figure}
\begin{center}
\includegraphics[width=7.5 cm]{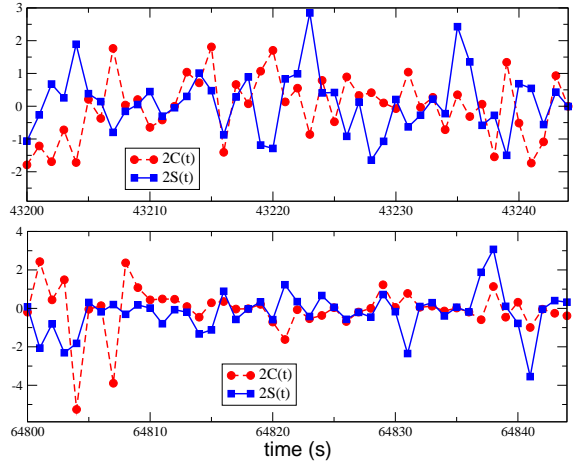}
\end{center}
\caption{For $\epsilon_v$ as in Eq.(\ref{refractive2}) and
$z=2$, we report in units $10^{-15}$ two typical sets of 45 seconds
for the two functions $2C(t)$ and $2S(t)$ of Eq.(\ref{basic3new}).
The two sets belong to the same random sequence and refer to two
sidereal times that differ by 6 hours. The boundaries of the
stochastic velocity components Eqs.(\ref{vx}) and (\ref{vy}) are
controlled by $(V,\alpha,\gamma)_{\rm CMB}$ through
Eqs.(\ref{projection}) and (\ref{isot}). For a laser frequency of
$2.8\cdot 10^{14}$ Hz \cite{schiller2015}, the range $\pm 3.5\cdot
10^{-15}$ corresponds to a typical frequency shift $\Delta \nu$ in
the range $\pm 1$ Hz, as in our
Fig.\ref{Figcrossed}.}
\label{rotation}
\end{figure}

The present analysis of the data is the following. For short-period
observations of a few days, the frequency shifts, measured upon
rotation of the apparatus, are used to extract the instantaneous
2C(t) and 2S(t) through Eq.(\ref{basic3new}). These data are then
compared with the parameterizations Eqs.(\ref{amorse1}) and
(\ref{amorse2}) to fit the $C_k$ and $S_k$ Fourier coefficients.
From very extensive observations, the present values of these
coefficients are at the level $10^{-18}\div10^{-19} $, i.e. about
1000 times smaller than the typical $10^{-15}$ instantaneous signal.

By recalling our discussion at the beginning of Sect.5, this is
exactly the same strategy traditionally adopted for the fringe
shifts in the old experiments and that cannot be maintained with a
genuine irregular signal. In fact, within our isotropic model, see
Eqs.(\ref{vanishing}) and (\ref{vanishing2})), one would find
$\langle C(t)\rangle_ {\rm stat}=0$ and $\langle S(t)\rangle_ {\rm
stat}=0$ at any time $t$ and mean values $(C_k)^{\rm avg}=0$,
$(S_k)^{\rm avg}=0$ for all Fourier coefficients. Therefore, with an
irregular but genuine signal a different type of analysis is needed.

To compare with the data, we have performed numerical simulations in
our isotropic stochastic model of Sect.4 with $\epsilon_v$ as in
Eq.(\ref{refractive2}). As a first illustration, we show in
Fig.\ref{rotation} two sequences of the instantaneous values for
2C(t) and 2S(t). The two sets belong to the same random sequence and
refer to two sidereal times that differ by 6 hours. The set
$(V,\alpha,\gamma)_{\rm CMB}$ was adopted to control the boundaries
of the stochastic velocity components through Eqs.(\ref{nassau1}),
(\ref{projection}) and (\ref{isot}). The value $\phi= 52$ degrees
was also fixed to reproduce the average latitude of the laboratories
in Berlin and D\"usseldorf. For a laser frequency of $2.8\cdot
10^{14}$ Hz \cite{schiller2015}, the interval $\pm 3.5\cdot
10^{-15}$ of these dimensionless amplitudes corresponds to a random
instantaneous frequency shift $\Delta \nu$ in the typical range $\pm
1$ Hz, as in our Fig.\ref{Figcrossed}.

For a more quantitative analysis we have considered the result of
ref.\cite{schiller2015} for the average variation of the frequency
shift over 1 second, see their Fig.3, bottom part. This corresponds
to a Root Square of the Allan Variance (RAV) of about $0.24$ Hz, or
$8.5 \cdot 10^{-16}$ at a fractional level. In general the RAV
describes the time dependence of an arbitrary function $f=f(t)$
which can be sampled over time intervals of length $\tau$. By
defining \BE {\overline f}(t_i;\tau)={{1}\over{\tau }}\int^{t_i+\tau
}_{t_i}dt~f(t)\equiv {\overline f}_i \EE one generates a
$\tau-$dependent distribution of ${\overline f}_i$ values. In a
large time interval $\Lambda= M\tau$, the RAV is then defined as \BE
\sigma_A(f,\tau)= \sqrt{\sigma^2_A(f,\tau) } \EE where \BE
\sigma^2_A(f,\tau)= {{1}\over{2(M-1) }}\sum^{M-1}_{i=1}
\left({\overline f}_i-{\overline f}_{i+1} \right)^2 \EE The
integration time $\tau$ is given in seconds and the factor of 2 is
introduced to obtain the same standard variance for uncorrelated
data as for a white-noise signal with uniform spectral amplitude at
all frequencies.

To understand the characteristics of our signal, we have thus
simulated one-day measurements of $2C(t)$ and $2S(t)$ at steps of 1
second. The RAV and the standard variance agree to good accuracy, so
that the signal of our isotropic stochastic model could be
approximated as a pure white noise. From these simulations of
one-day measurements, ($z=$ 1 or 2), we obtained mean values
$\langle 2C\rangle_{\rm day} =-1.6 \cdot(z/2) \cdot 10^{-18}$,
$\langle 2S\rangle_{\rm day} =4.3 \cdot (z/2) \cdot 10^{-18}$ and
variances
\begin{equation} \label{sigmac}
\left[ \sigma_A(2C,1) \right]_{\rm simul } = {{z}\over{2}}(8.7 \pm
0.8)\cdot 10^{-16} \EE \BE \label{sigmas}\left[ \sigma_A(2S,1)
\right]_{\rm simul}= {{z}\over{2}}(9.6 \pm 0.9)\cdot 10^{-16}
\end{equation} Here the $\pm$ uncertainties reflect the observed variations
due to the truncation of the Fourier modes in Eqs.(\ref{vx}),
(\ref{vy}) and to the dependence on the random sequence. From
Eq.(\ref{basic3new}), by combining quadratically these two sigma's,
we estimate
\begin{equation}
\left[\sigma_A({{\Delta \nu }\over{\nu_0}},1)\right]_{\rm simul
}\sim  \sqrt {{{1}\over{2}}~\sigma^2_A[2C,1]_{\rm simul}  +
{{1}\over{2}}~\sigma^2_A[2S,1]_{\rm simul} } \sim {{z}\over{2}}
(9.2\pm 0.9) \cdot 10^{-16}\end{equation} so that, for a laser
frequency $\nu_0=2.8\cdot 10^{14}$ Hz \cite{schiller2015}, we would
predict an average RAV
\begin{equation}
 \label{ourallan}
\left[ \sigma_A(\Delta \nu,1) \right]_{\rm simul } \sim
{{z}\over{2}}(0.26 \pm 0.03) ~{\rm Hz}\end{equation} of the
frequency shift at 1 second. This estimate should be compared with
the mentioned experimental value
\begin{equation} \left[ \sigma_A(\Delta \nu,1) \right]_{\rm
exp}\sim 0.24~ {\rm Hz}
\end{equation} reported in ref.\cite{schiller2015}.
The good agreement with our simulated value indicates that, at least
for an integration time of 1 second, the correction to our model
should be negligible. Also, the data favor $z=2$, which is the only
free parameter of our scheme.

Our model, however, makes another definite prediction: during the
day there should be characteristic modulations which reflect the
periodic variations of  $\tilde v(t)$ Eq.(\ref{projection}) in the
plane of the interferometer. For $z=2$ and the typical
Central-Europe value $\tilde{v}(t)=(250\div 370)$ km/s, taking into
account uncertainties in the simulations, from bins of data centered
around the various times $t$, the RAV at 1 second explores the range
\begin{equation} \label{bounds} 5 \cdot 10^{-16} \lesssim~
\left[\sigma_A({{\Delta \nu }\over{\nu_0}},1)\right]_t~ \lesssim
12\cdot 10^{-16}\end{equation} This range is obtained with our
numerical simulation but can be approximated as
\begin{equation} \label{boundssimple}
\left[\sigma_A({{\Delta \nu }\over{\nu_0}},1)\right]_t\sim ~
8.4\cdot 10^{-16} \left({{\tilde v(t) }\over{315~ {\rm km/s}
}}\right)^2\end{equation} where $\tilde v(t)$ is defined in
Eq.(\ref{projection}).

Detecting these periodic variations would therefore give the
cleanest test of our picture, provided these variations are not
obscured by spurious effects. The simplest strategy for a comparison
is to determine, from the spectral amplitude of the experimental
signal, the frequency $\omega_0$ beyond which the spectral amplitude
$\sqrt{S(\omega)}$ becomes flat. Thus, by defining $\tau_0\sim
\omega_0^{-1}$, for integration times $\tau\lesssim \tau_0$ the RAV
is dominated by the pure white-noise component of the signal. Then,
since typically $\tau_0\sim$ 1 second, by measuring the experimental
RAV at $\tau_0$, in different hours of the day, one can directly
compare with Eq.(\ref{bounds}) \footnote{However, the time $\tau_0$
could also be considerably larger than 1 second as, for instance, in
the cryogenic experiment of ref.\cite{mueller2003}. There, the RAV
at 1 second was about 10 times larger than the range
Eq.(\ref{bounds}) but, in the quiet phase between two refills of the
refrigerator, $\sigma_A(\Delta \nu/\nu_0,\tau)$ was monotonically
decreasing as $\tau^{-1/2}$ up to $\tau_0=250$ seconds where it
reached its minimum value $\sigma_A(\Delta \nu/\nu_0,\tau_0)\sim
5.3\cdot 10^{-16}$. This is still consistent with the lower bound in
Eq.(\ref{bounds}) so that we would tentatively argue that
Eq.(\ref{bounds}) should be replaced by the more general form
$5\cdot 10^{-16} \lesssim~ [\sigma_A(\Delta \nu/\nu_0),\tau_0)]_t~
\lesssim 12\cdot 10^{-16}$, with the same range but a $\tau_0$ which
now depends on the experiment.}.

For a more refined comparison, one could try to generate a colored
signal which, as in the real experimental situation, contains
various branches (white-noise, pink-noise, random-walk...), and
estimate directly the modifications of our basic white-noise
component at the various $\tau$'s. Since these modifications depend
on the particular experiment, we have decided to consider
ref.\cite{nagelnature}. This is a high-precision cryogenic
experiment, with microwaves of 12.97 GHz, where almost all
electromagnetic energy propagates in a medium, sapphire, with
refractive index of about 3 (at microwave frequencies). Therefore,
an analysis of this experiment will also check our
Eq.(\ref{anysolid}) implying that a fundamental $10^{-15}$ vacuum
signal as in  (\ref{6introany}), with very precise measurements,
should also show up in a solid dielectric.

\begin{figure}
\begin{center}
\includegraphics[width=8.5 cm]{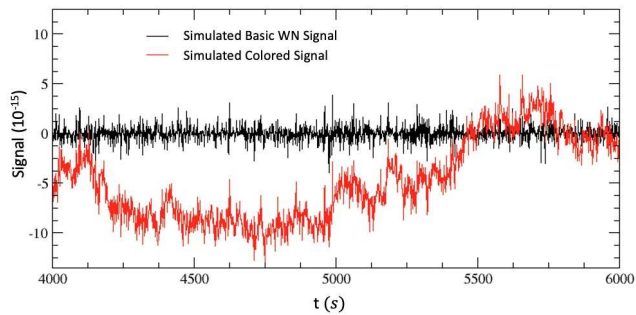}
\end{center}
\caption{We report two typical sets of 2000 seconds for our
basic white-noise (WN) signal and its colored version obtained by
Fourier transforming the spectral amplitude of
ref.\cite{nagelnature}.}
\label{colored}
\end{figure}

From Figure 3(c) of \cite{nagelnature}, the spectral amplitude of
this particular apparatus is seen to become flat at frequencies
$\omega \ge 0.5$ Hz indicating the order of magnitude estimate
$\tau_0\sim$ 1 second. In collaboration with Dr. Giancarlo Cella of
the VIRGO Collaboration, these data for the spectral amplitude were
then fitted to an analytic, power-law form to describe the
lower-frequency part 0.001 Hz $\leq \omega \leq 0.5$ Hz. This fitted
spectrum was then used to generate a signal by Fourier transform.
Finally, very long sequences of this signal were stored to produce
``colored'' version of our basic white-noise signal. The details of
this analysis will be published elsewhere \cite{cella}.

\begin{figure}
\begin{center}
\includegraphics[width=9.5 cm]{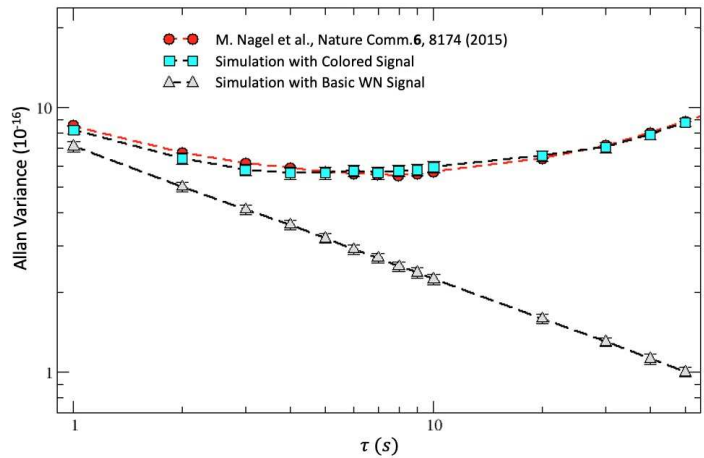}
\end{center}
\caption{We report the Allan variance for the fractional
frequency shift obtained from simulations of sequences of 2000
seconds for our basic white-noise (WN) signal and for its colored
version obtained by Fourier transforming the spectral amplitude of
ref.\cite{nagelnature}. The direct experimental results of
ref.\cite{nagelnature}, for the non-rotating setup, are also
shown.}
\label{allan}
\end{figure}

Here we will limit ourselves to report the results of a first set of
simulations in intervals of 2000 seconds. To get a qualitative
impression of the effect, we report in Fig.\ref{colored} a sequence
of our basic white-noise signal and a sequence of its colored
version. By averaging over many 2000-second sequences of this type,
the corresponding RAV's for the two signals are reported in
Fig.\ref{allan}. The experimental RAV extracted from Figure 3(b) of
ref.\cite{nagelnature} is also reported (for the non-rotating
setup). At this stage, the agreement of our simulated, colored
signal with the experimental data remains satisfactory only up
$\tau=$ 50 seconds. Reproducing the signal at larger $\tau$'s will
require further efforts but this is not relevant here, our scope
being just to understand the modifications of our stochastic signal
near the 1-second scale.

From Fig.\ref{allan} we find that, at the value of interest $\tau=$
1 second, our predicted white-noise signal $ (7.1 \pm 0.3)\cdot
10^{-16}$ is changed respectively by about $+15\%$, when comparing
with our simulated colored value $ (8.2 \pm 0.3)\cdot 10^{-16}$, or
by about $+20\%$, when comparing with the experimental value of
about $ 8.5 \cdot 10^{-16}$. Thus periodic variations of a factor of
2 as in Eq.(\ref{bounds}), if present in the experimental data,
should remain visible, at least with a systematics at the level of
ref.\cite{nagelnature}.

At the same time, this $ 8.5 \cdot 10^{-16}$, obtained in
ref.\cite{nagelnature} for the experimental RAV at 1 second, is the
same $ 8.5 \cdot 10^{-16}$ that we extracted from the value
$\sigma_A(\Delta \nu,1)_{\rm exp}\sim $ 0.24 Hz of
ref.\cite{schiller2015} after normalizing to the laser frequency
$\nu_0=2.8 \cdot 10^{14}$ Hz. Therefore this beautiful agreement,
between ref.\cite{schiller2015} (a vacuum experiment at room
temperature) and ref.\cite{nagelnature} (a cryogenic experiment in a
solid dielectric), while confirming our predictions Eqs.
(\ref{6introany}) (\ref{anysolid}) of a fundamental $10^{-15}$
signal, indicates that periodic variations as in Eq.(\ref{bounds})
should also remain visible with the apparatus of
ref.\cite{schiller2015}.

\section{Summary and conclusions}

Due to the present interpretation of the dominant dipole anisotropy
of the Cosmic Microwave Background as a Doppler effect, one may
wonder about the reference system where this dipole vanishes
exactly. Since the observed motion is, to good approximation, the
combination of peculiar motions and reflects local inhomogeneities,
one could naturally consider the idea of a global frame of rest,
associated with the Universe as a whole, which could characterize
the form of relativity physically realized in nature. The isotropy
of the CMB could then just {\it indicate} the existence of this
fundamental system $\Sigma$ that we could conventionally decide to
call ``ether'' but the cosmic radiation itself would not {\it
coincide} with this type of ether. Due to the fundamental group
properties of Lorentz transformations, two observers, individually
moving with respect to $\Sigma$, would still be connected by the
standard relativistic composition rule of velocities. But ultimate
implications could be far reaching. Just think about the
interpretation of non-locality in the quantum theory.

Since the answer cannot be found on a pure theoretical ground,
physical interpretation is traditionally postponed to the detection
of some dragging of light in the earth frame. Namely, to measuring a
small angular dependence ${{\Delta\bar{c}_\theta}\over{c}}$ of the
two-way velocity of light in laboratory and trying to correlate the
measurements with the direct CMB observations with satellites in
space.  The present view is that no such meaningful correlation has
ever been observed, all data collected so far (from Michelson-Morley
to the modern experiments with optical resonators) being just
considered typical instrumental effects in measurements with better
and better systematics.

However, if the velocity of light in the interferometers is not the
same parameter ``c'' of Lorentz transformations, nothing would
prevent a non-zero dragging. For instance, in experiments in gaseous
media with refractive index ${\cal N}=1 + \epsilon$, the small
fraction of refracted light could keep track of the velocity of
matter with respect to the hypothetical $\Sigma$ and produce a
direction-dependent refractive index. Then, from symmetry arguments
valid in the $\epsilon \to 0$ limit, one would expect
${{|\Delta\bar{c}_\theta|}\over{c}} \sim \epsilon (v^2/c^2)$ which
is much smaller than the classical expectation
${{|\Delta\bar{c}_\theta|_{\rm class}}\over{c}} \sim (v^2/2c^2)$.
For $v\sim$ 300 km/s, and inserting the appropriate refractive
index, i.e. $\epsilon\sim 2.8\cdot10^{-4}$ for air and $\epsilon\sim
3.3 \cdot10^{-5}$ for gaseous helium, this reproduces the observed
order of magnitude, respectively ${{|\Delta\bar{c}_\theta|}\over{c}}
\sim 10^{-10}$ and ${{|\Delta\bar{c}_\theta|}\over{c}} \sim
10^{-11}$.

In addition, besides being much smaller than classically expected,
observable effects could have an irregular nature. This means that
the projection of the global velocity field at the site of the
experiment, say $ \tilde v_\mu(t)$, could differ non trivially from
the local field $v_\mu(t)$ which determines the instantaneous
direction and magnitude of the drift in the plane of the
interferometer. As a definite model, to relate $v_\mu(t)$ and  $
\tilde v_\mu(t)$, on the basis of some theoretical arguments, we
have followed the physical analogy with a turbulent fluid, in
particular, with that form of turbulence which, at small scales,
becomes statistically homogeneous and isotropic. To this end, the
local $v_\mu(t)$ was expanded in a large number of Fourier
components varying randomly within boundaries which depend on the
smooth $\tilde{v}_\mu(t)$ determined by the average motion of the
earth. In this model, at the small scale of the experiment,
statistical averages of vector quantities vanish identically.
Therefore, one should analyze the data for
${{\Delta\bar{c}_\theta(t)}\over{c}}$ in phase $\theta_2(t)$ and
amplitude $A_2(t)$, which give respectively the direction and
magnitude of the local drift, and concentrate on the latter which,
being positive definite, remains non-zero under any averaging
procedure. Then, even discarding $\theta_2(t)$, the time modulations
of the statistical average $\langle A_2(t) \rangle_{\rm stat}$ could
still be used to correlate a genuine signal with the corresponding
cosmic motion.

As a proof, we report some remarkable correlations found in the old
experiments:

a) by fitting with Eqs.(29) and (30) the smooth polynomial
interpolation of the irregular Joos 2nd-harmonic amplitudes in our
Fig.7, one finds \cite{plus} a right ascension $\alpha({\rm
fit-Joos})= (168 \pm 30)$ degrees and an angular declination
$\gamma({\rm fit-Joos})= (-13 \pm 14)$ degrees which are well
consistent with the present values $\alpha({\rm CMB}) \sim$ 168
degrees and $\gamma({\rm CMB}) \sim -$7 degrees.

b) by inspection of our Table 2, if we compare with our
Eq.(\ref{amplitude10001final}), all experiments with light
propagating in air give $\tilde v_{\rm air} \sim 418 \pm 62 $ km/s
and the two experiments in gaseous helium $\tilde v_{\rm helium}
\sim 323 \pm 70 $ km/s. Thus the global average $\langle\tilde v
\rangle\sim 376 \pm 46 $ km/s agrees well with the 370 km/s from the
direct CMB observations.

c) from the two most precise experiments in Table 2, Piccard-Stahel
(Brussels and Mt. Rigi in Switzerland) and Joos (Jena), we find two
determinations, $\tilde v= 360^{+85}_{-110} $ km/s and $\tilde v=
305^{+85}_{-100} $ km/s respectively, whose average $\langle \tilde
v\rangle \sim 332^{+60}_{-80} $ km/s reproduces to high accuracy the
projection of the CMB velocity at a typical Central-Europe latitude.

Still the simple relation ${{|\Delta\bar{c}_\theta(t)|}\over{c}}\sim
\epsilon (v^2(t)/c^2)$, while providing a consistent description,
does not explain how the earth motion produces the observed small
anisotropy in the gaseous systems. Here, in this summary, rather
than re-proposing immediately our reasoning of Sect.6, we shall
follow the other way round. We will thus first summarize the
analysis of Sect.7, for the present experiments in vacuum and in
solid dielectrics, and at the very end, armed with these results,
return to the mechanism at work in the gaseous media.

In Sect.7, we started from the modern experiments which measure the
frequency shift $\Delta \nu(t)$ of two vacuum optical resonators. By
considering the most precise experiments, with optical cavities made
of different materials, and operating at room temperature and in the
cryogenic regime, one gets the idea of a universal, irregular signal
with typical fractional magnitude ${{|\Delta \nu(t)| }\over{\nu_0}}
\sim 10^{-15}$. Within the same model adopted for the classical
experiments, we have thus explored the possibility to interpret this
signal in terms of a vacuum refractivity $\epsilon_v ={\cal N}_v -
1\sim 10^{-9}$ in order to obtain ${{|\Delta \nu(t)| }\over{\nu_0}}
\sim \epsilon_v (v(t)/c)^2 \sim 10^{-15}$ for the typical $v(t)\sim$
300 km/s.

This $10^{-9}$ vacuum refractivity could have a precise physical
interpretation. In fact, the value $\epsilon_v \sim (2G_N M/c^2
R)\sim 1.4 \cdot 10^{-9}$ was suggested \cite{gerg} as a possible
signature to distinguish an apparatus on the earth surface from the
same apparatus placed in that ideal freely-falling frame which
defines the parameter $c$ of Lorentz transformations, see
Fig.\ref{freefall}. In addition, in our stochastic model, a definite
$10^{-15}$ instantaneous signal will coexist with vanishing
statistical averages for all vector quantities, such as the $C_k$
and $S_k$ Fourier coefficients extracted from a standard temporal
fit to the data with Eqs.(\ref{amorse1}) and (\ref{amorse2}). Our
physical model, would thus be immediately consistent with the
present $10^{-18}\div 10^{-19}$ limits obtained for these
coefficients after averaging over many observations.

Since our signal can be approximated as a universal form of white
noise and sets an intrinsic limit to the precision of measurements,
for a comparison with experiments, we have then considered the
characteristics of the signal for that integration time (typically 1
second) where the pure white-noise branch is as small as possible
but other types of noise are not yet important. In this case, when
comparing with ref.\cite{schiller2015}, our numerical simulation of
the Allan variance for measurements during a whole day,
$\sigma_A(\Delta \nu,1)_{\rm simul}= 0.26 \pm 0.03 $ Hz is in
complete agreement with the experimental result $\sigma_A(\Delta
\nu,1)_{\rm exp}\sim $ 0.24 Hz.

We have also emphasized that this 0.24 Hz, when normalized to their
laser frequency, gives a fractional shift of $ 8.5 \cdot 10^{-16}$
which is precisely the same obtained, at 1 second, in
ref.\cite{nagelnature}, see our Fig.\ref{allan}. Now,
ref.\cite{schiller2015} is an experiment running with vacuum
cavities, at room temperature and with reference frequency
$\nu_0=2.8 \cdot 10^{14}$ Hz. While ref.\cite{nagelnature} is a
cryogenic experiment, with microwaves of 12.97 GHz, where almost all
electromagnetic energy propagates in a medium, sapphire, with
refractive index of about 3. It is impossible that this
extraordinary agreement can be due to accidental effects. Therefore,
our conclusion: there is a fundamental vacuum signal which shows up
in vacuum and in solid dielectrics and whose average magnitude is
completely consistent with the vacuum refractivity of
ref.\cite{gerg} and the velocity of 370 km/s from the CMB
observation with satellites in space.

We also predict periodic, daily variations in the range $(5\div12)
\cdot 10^{-16}$ for a typical Central-Europe latitude. This range
was obtained from our numerical simulation but can also be expressed
in a simpler way as
\begin{equation} \label{boundssimple2}
\left[\sigma_A({{\Delta \nu }\over{\nu_0}},1)\right]_t\sim ~
8.4\cdot 10^{-16} \left({{\tilde v(t) }\over{315~ {\rm km/s}
}}\right)^2\end{equation} where $\tilde v(t)$ is defined in
Eq.(\ref{projection}). Our simulations at the end of Sect.7 indicate
that, for integration time of 1 second, our basic signal is modified
by about $20\%$. Therefore these periodic variations, if there,
should remain visible.

Let us then return to gaseous media. Namely, which could be a
physical mechanism that, starting from a fundamental $10^{-15}$
vacuum signal, enhances the effect, respectively up to $10^{-11}$
and $10^{-10}$ in gaseous helium and air, and finally disappears in
solid dielectrics, as in the mentioned very precise cryogenic
experiment in sapphire, which gives again the same $10^{-15}$ as in
vacuum? Our answer to this question, in Sect.6, was based on the
traditional interpretation \cite{joos2,shankland} of those old
residuals in terms of small temperature difference $\Delta T^{\rm
gas}(\theta)$, of a millikelvin or so, in the gas of the two optical
arms. We have, however, obtained the same universal value $\Delta
T^{\rm gas}(\theta)= 0.2\div 0.3$ mK from the various experiments.
Therefore those old estimates, besides being slightly too large,
were not catching the basic point: different experiments converge
toward the same value and, therefore, the thermal effect cannot be
due to local temperature conditions but must have a {\it non-local}
origin. Our interpretation is that the interactions of the gas
molecules with the background radiation are so weak that, on
average, only less than 1/10 of the $\Delta T^{\rm CMB}(\theta)$ in
Eq.(\ref{CBR}) is transferred to bring the gas out of equilibrium.
Nevertheless, regardless of its precise value, a universal $\Delta
T^{\rm gas}(\theta)\lesssim $ 1 mK can help intuition by explaining
the quantitative reduction of the effect in the vacuum limit, where
$\epsilon_{\rm gas} \to 0$, and the qualitative difference with
solid dielectrics where such tiny temperature differences become
irrelevant. We have also observed that, after a century from those
old experiments, in room-temperature measurements, values $\Delta
T\lesssim $ 1 mK are still state of the art for the precision
attainable in temperature differences, see
e.g.\cite{farkas,zhaoa,trusov}.

In conclusion, by considering old and modern experiments, we have
found several correlations between optical measurements in
laboratory and the kinematical parameters obtained from the direct
CMB observations with satellites in space. These correlations are
summarized in the three items a), b) and c) listed above and in the
successful quantitative description of the RAV measured in
refs.\cite{schiller2015} and \cite{nagelnature} for the relevant
region of integration times of about 1 second where the white-noise
branch is as small as possible but other experiment-dependent
effects are not yet important. Ours is not the only scheme to
analyze the experiments but, yet, fulfills the criterion
traditionally adopted to indicate a reference system which could
play the role of fundamental frame for relativity. We also observe
that, for the same region of integration times, our scheme predicts
periodic daily variations of the RAV which should be observable.
Therefore, for the importance of the issue, we would expect to
receive an experimental confirmation or a disproof. If definitely
confirmed, one more complementary test should be performed by
placing the vacuum (or solid dielectric) optical cavities on board
of a satellite, as in the OPTIS proposal \cite{optis}. In this ideal
free-fall environment, as in panel (a) of our Fig.\ref{freefall},
the typical instantaneous frequency shift should be much smaller (by
orders of magnitude) than the corresponding $10^{-15}$ value
measured with the same interferometers on the earth surface.

\centerline{\bf Acknowledgments}

\par\noindent We thank Giancarlo Cella for useful discussions and his
collaboration.


\begin{thebibliography}{99}

\bibitem{penzias}
A. A. Penzias and R. W. Wilson, Astrophys. J. {\bf 142}, 419 (1965).
\bibitem{partridge}
R. B. Partridge and D. T. Wilkinson, Phys. Rev. Lett. {\bf 18}, 557
(1967).
\bibitem{heer}
C. V. Heer, Phys. Rev. {\bf 174}, 1611 (1968).
\bibitem{mather}
J. C. Mather, Rev. Mod. Phys. {\bf 79}, 1331 (2007).
\bibitem{smoot}
G. F. Smoot, Rev. Mod. Phys. {\bf 79}, 1349 (2007).
\bibitem{yoon}
M. Yoon and D. Huterer, Ap. J. Lett. {\bf 813}, L18 (2015).
\bibitem{bell}
J. S. Bell, How to teach special relativity, in Speakable and
unspeakable in quantum mechanics, Cambridge University Press 1987,
pag. 67.
\bibitem{brownbook}
H. R. Brown, Physical Relativity. Space-time structure from  a
dynamical perspective, Clarendon Press, Oxford 2005.
\bibitem{guerraejtp}
R. de Abreu and V. Guerra, Electr. J. of Theor. Phys. {\bf 12}, 183
(2015).
\bibitem{shanahan}
D. Shanahan, Found. of Phys. {\bf 44}, 349 (2014).
\bibitem{ungar}
A. Ungar, Found. of Phys. {\bf 30}, 331 (2000).
\bibitem{costella}
J. P. Costella et al., Am. J. Phys. {\bf 69}, 837 (2001).
\bibitem{kanevisser}
K. O' Donnell and M. Visser, Eur. J. Phys. {\bf 32}, 1033 (2011).
\bibitem{hardy}
L. Hardy, Phys. Rev. Lett. {\bf 68} (1992) 2981.
\bibitem{thooft}
G. t Hooft, Search of the Ultimate Building Blocks, Cambridge Univ.
Press 1997, p.70.
\bibitem{mech}
M.~Consoli, P.M. Stevenson, Int. J. Mod. Phys. A \textbf{15}, 133
(2000).
\bibitem{epjc}
M. Consoli and E. Costanzo,  Eur. Phys. Journ. {\bf C54}, 585 (2008)
285.
\bibitem{dedicated}
M. Consoli and E. Costanzo, Eur. Phys. Journ. {\bf C55}, 469 (2008).
\bibitem{foop}
M. Consoli, Found. of Phys. {\bf 45}, 22 (2015).
\bibitem{Rubakov}
V.Rubakov, Phys. Usp. {\bf 51}, 759 (2008).
\bibitem{arrault}
I. Arraut, Europhysics Letters {\bf 111}, 61001 (2015).
\bibitem{deser}
S. Deser and R. P.Woodard, Phys.Rev.Lett. {\bf 99}, 111301 (2007).
\bibitem{soussa}
M. E. Soussa, R. P.Woodard, Class.Quant.Grav. {\bf 20}, 2737 (2003).
\bibitem{Nojiri}
S. Nojiri and S.D. Odintsov, Phys. Lett. B {\bf 659}, 821 (2008).
\bibitem{nagelnature}
M. Nagel  et al., Nature Comm.{\bf 6}, 8174 (2015).
\bibitem{mm}
A. A. Michelson and E. W. Morley, Am. J. Sci. {\bf 34}, 333 (1887).
\bibitem{miller}
D. C. Miller, Rev. Mod. Phys. {\bf 5}, 203 (1933).
\bibitem{kenconference}
A. A. Michelson, et al., Ap. J. {\bf 68} (1928) p. 341-402.
\bibitem{illingworth}
K. K. Illingworth, Phys. Rev. {\bf 30}, 692 (1927).
\bibitem{tomaschek1}
R. Tomaschek, Astron. Nachrichten, {\bf 219}, 301 (1923), English
translation.
\bibitem{piccard3}
A. Piccard and E. Stahel, Journ. de Physique  et Le Radium {\bf IX}
(1928) No.2.
\bibitem{mpp}
A. A. Michelson, F. G. Pease and F. Pearson, Nature, {\bf 123}, 88
(1929).
\bibitem{mpp2}
A. A. Michelson, F. G. Pease and F. Pearson, J. Opt. Soc. Am. {\bf
18}, 181 (1929).
\bibitem{pease}
F. G. Pease, Publ. of the Astr. Soc. of the Pacific, {\bf XLII}, 197
(1930).
\bibitem{joos}
G. Joos, Ann. d. Physik {\bf 7}, 385 (1930).
\bibitem{pla}
M. Consoli and E. Costanzo, Phys. Lett. A {\bf 333}, 355 (2004).
\bibitem{plus}
M. Consoli, C. Matheson and A. Pluchino, Eur. Phys. J. Plus {\bf
128}, 71 (2013).
\bibitem{epl}
M. Consoli, A. Pluchino and A. Rapisarda:  Europhysics Lett. {\bf
113}, 19001 (2016).
\bibitem{plus2}
M. Consoli and A. Pluchino, Eur. Phys. Jour. Plus {\bf 133}, 295
(2018).
\bibitem{book}
M. Consoli and A. Pluchino, Michelson-Morley Experiments: an Enigma
for Physics and the History of Science, World Scientific 2019, ISBN
978-981-3278-18-9.
\bibitem{chaos}
M. Consoli, A. Pluchino and A. Rapisarda,  Chaos, Solitons and
Fractals {\bf 44}, 1089 (2011).
\bibitem{physica}
M. Consoli, A. Pluchino, A. Rapisarda and S. Tudisco, Physica {\bf
A394}, 61 (2014).
\bibitem{fox}
J. Shamir and R. Fox, N. Cim. {\bf 62B}, 258 (1969).
\bibitem{joos2}
G. Joos, Phys. Rev. {\bf 45}, 114 (1934).
\bibitem{shankland}
R. S. Shankland et al., Rev. Mod. Phys.{\bf 27}, 167 (1955).
\bibitem{mueller2003}
H. M\"uller, et al. , Phys. Rev. Lett. {\bf 91}, 020401 (2003).
\bibitem{crossed}
Ch. Eisele et al.,  Opt. Comm. {\bf 281}, 1189 (2008).
\bibitem{newberlin}
S. Herrmann, et al., Phys.Rev. D {\bf 80}, 10511 (2009).
\bibitem{newschiller}
Ch. Eisele, A. Newsky  and S. Schiller, Phys. Rev. Lett. {\bf 103},
090401 (2009).
\bibitem{cpt2013}
M. Nagel et al., Ultra-stable Cryogenic Optical Resonators For Tests
Of Fundamental Physics, arXiv:1308.5582[physics.optics].
\bibitem{schiller2015}
Q. Chen, E. Magoulakis, and S. Schiller,Phys. Rev. D {\bf 93 },
022003 (2016).
\bibitem{gerg}
M. Consoli and L. Pappalardo, Gen. Rel. and Grav. {\bf 42}, 2585
(2010).
\bibitem{pound}
R. V. Pound, Rev. Sci. Instrum. {\bf 17}, 490 (1946).
\bibitem{PDH}
R. W. P. Drever et al., Appl. Phys. B {\bf 31}, 97 (1983).
\bibitem{black}
E. D. Black, Am. J. Phys. {\bf 69}, 79 (2001).
\bibitem{guerra2005}
V. Guerra and R. de Abreu, Eur. J. of Phys. {\bf 26}, S117 (2005).
\bibitem{maxwell}
J. C. Maxwell, Ether, Encyclopaedia Britannica, 9th Edition, 1878.
\bibitem{leonhardt}
U. Leonhardt and P. Piwnicki, Phys. Rev. {\bf A60}, 4301 (1999).
\bibitem{jauch}
J. M. Jauch and K. M. Watson, Phys. Rev. {\bf 74}, 950 (1948).
\bibitem{kennedy}
R. J. Kennedy, Phys. Rev. {\bf 47}, 965 (1935).
\bibitem{feybook}
R. P. Feynman, R. B. Leighton and M. Sands, The Feynman Lectures on
Physics, Addison Wesley Publ. Co. 1963.
\bibitem{onsager}
L. Onsager, Nuovo Cimento, Suppl. {\bf 6}, 279 (1949).
\bibitem{eyink}
G. L. Eyink and K. R. Sreenivasan Rev. Mod. Phys. {\bf 78}, 87
(2006).
\bibitem{nassau}
J. J. Nassau and P. M. Morse, Ap. J. {\bf 65}, 73 (1927).
\bibitem{landau}
L. D. Landau and E. M. Lifshitz, Fluid Mechanics, Pergamon Press
1959, Chapt. III.
\bibitem{fung}
J. C. H. Fung et al., J. Fluid Mech. {\bf 236}, 281 (1992).
\bibitem{miller34}
D. C. Miller, Phys. Rev. {\bf 45} (1934) 114.
\bibitem{swensonbook}
L. S. Swenson Jr., the Ethereal Aether, A History of the
Michelson-Morley-Miller Aether-Drift Experiments, 1880-1930.
University of Texas Press, Austin 1972.
\bibitem{loyd2}
Loyd S. Swenson Jr., Journ. for the History of Astronomy, {\bf 1},
56 (1970).
\bibitem{stone}
J. A. Stone and A. Stejskal, Metrologia {\bf 41}, 189 (2004).
\bibitem{jaseja}
T. S. Jaseja, et al.,  Phys. Rev. {\bf 133}, A1221 (1964).
\bibitem{farkas}
E. R. Farkas and W. W. Webb, Rev. Scient. Instr. 81, 093704 (2010).
\bibitem{zhaoa}
Y. Zhaoa, D. L. Trumperb, R. K. Heilmann, M. L. Schattenburg,
Precision Engin. 34, 164 (2010).
\bibitem{trusov}
I. P. Prikhodko, A. A. Trusov, A. M. Shkel, Sensors and Actuators A
201, 517 (2013).
\bibitem{nist}
D. A. Jennings et al. , Journ. of Res. Nat. Bur. Stand. {\bf 92}, 11
(1987).
\bibitem{brillet}
A. Brillet and J. L. Hall, Phys. Rev. Lett. {\bf 42}, 549 (1979).
\bibitem{numata}
K. Numata, A, Kemery and J. Camp, Phys. Rev. Lett. {\bf 93}, 250602
(2004).
\bibitem{barcelo1}
C. Barcelo, S. Liberati and M. Visser, Class. Quantum Grav. {\bf
18}, 3595 (2001).
\bibitem{barcelo2}
M. Visser, C. Barcelo and S. Liberati, Gen. Rel. Grav. {\bf 34},
1719 (2002).
\bibitem{volo} G. E. Volovik, Phys. Rep. {\bf 351}, 195
(2001).
\bibitem{bosegravity}
R. Sch\"utzhold, Class. Quantum Gravity {\bf 25}, 114027 (2008).
\bibitem{ultraweak}
M. Consoli, Class. Quantum Grav. {\bf 26}, 225008 (2009).
\bibitem{jannes}
G. Jannes and G. E. Volovik,  JETP Lett.{\bf 96}, 215 (2012).
\bibitem{cosmo}
S. Finazzi, S. Liberati and L. Sindoni,  Phys. Rev. Lett. {\bf 108},
071101 (2012).
\bibitem{yilmaz}
H. Yilmaz, Phys. Rev. {\bf 111}, 1417 (1958).
\bibitem{tupper}
B. O. J. Tupper, N. Cimento {\bf 19B}, 1974 (135); Lett. N. Cimento
{\bf 14}, 627 (1974).
\bibitem{rule}
R. P. Feynman, in Superstrings: A Theory of Everything ?, P. C. W.
Davies and J. Brown Eds., Cambridge University Press, 1997, pag.
201.
\bibitem{cook}
R. J. Cook, Am. J. Phys. {\bf 72}, 214 (2004).
\bibitem{broekaert}
J. Broekaert, Found. of Phys. {\bf 38}, 409 (2008).
\bibitem{cella}
G. Cella, M. Consoli and A. Pluchino, in preparation.
\bibitem{optis}
C. L\"ammerzahl et al., Class. Quantum Gravity {\bf 18}, 2499
(2001).



\end{thebibliography}
\end{document}